\documentclass[traditabstract]{aa}
\usepackage{graphicx}
\usepackage{txfonts}
\usepackage{longtable}
\usepackage{xcolor}
\usepackage{natbib}
\usepackage{pdfpages}
\usepackage{verbatim}

\bibpunct{(}{)}{;}{a}{}{,}
\begin{document} 

%%%%%%%%%%%%%%%%%%%%%%%%%%%%
%New commands
%%%%%%%%%%%%%%%%%%%%%%%%%%%%
\newcommand{\hi}{\mbox{H\,{\sc i}}}
\newcommand{\zabs}{$z_{\rm abs}$}
\newcommand{\zmin}{$z_{\rm min}$}
\newcommand{\zmax}{$z_{\rm max}$}
\newcommand{\zq}{$z_{\rm q}$}
\newcommand{\zg}{$z_{\rm g}$}
\newcommand{\kms}{km\,s$^{-1}$}
\newcommand{\cmsq}{cm$^{-2}$}
\newcommand{\degree}{\ensuremath{^\circ}}
\newcommand{\Msun}{$M_{\odot}$} 
\newcommand{\mgii}{\mbox{Mg\,{\sc ii}}} 
\newcommand{\mgiia}{\mbox{Mg\,{\sc ii}$\lambda$2796}}
\newcommand{\mgiib}{\mbox{Mg\,{\sc ii}$\lambda$2803}}
\newcommand{\mgiiab}{\mbox{Mg\,{\sc ii}$\lambda\lambda$2796,2803}}
%%%%
\newcommand{\lapp}{\mbox{\raisebox{-0.3em}{$\stackrel{\textstyle <}{\sim}$}}}
\newcommand{\gapp}{\mbox{\raisebox{-0.3em}{$\stackrel{\textstyle >}{\sim}$}}}
\newcommand{\pks}{PKS\,1830$-$211}

%%%%%%%%%%%%%%%%%%%%%%%%%%%%%%%%%%%
%Institutes
\newcommand{\lerma}{Observatoire de Paris, LERMA, Coll\`ege de France, CNRS, PSL University, Sorbonne University, 75014, Paris --- \email{francoise.combes@obspm.fr} \label{lerma}}
\newcommand{\iucaa}{Inter-University Centre for Astronomy and Astrophysics, Post Bag 4, Ganeshkhind, Pune 411 007, India \label{iucaa}}
\newcommand{\iap}{Institut d'Astrophysique de Paris, CNRS-SU, UMR\,7095, 98bis bd Arago, 75014 Paris, France \label{iap}}
\newcommand{\ioffe}{Ioffe Institute, {Polyteknicheskaya 26}, 194021 Saint-Petersburg, Russia \label{ioffe}}
\newcommand{\chalmers}{Department of Space, Earth and Environment, Chalmers University of Technology, Onsala Space Observatory, 43992 Onsala, Sweden \label{chalmers}}
\newcommand{\sarao}{South African Radio Astronomy Observatory, 2 Fir Street, Black River Park, Observatory 7925, South Africa \label{sarao}}
\newcommand{\rhodes}{Department of Physics and Electronics, Rhodes University, P.O. Box 94, Makhanda, 6140, South Africa \label{rhodes}}
\newcommand{\argl}{Argelander-Institut f\"ur Astronomie, Auf dem H\"ugel 71, D-53121 Bonn, Germany \label{argl}}
\newcommand{\nrao}{National Radio Astronomy Observatory, Socorro, NM 87801, USA \label{nrao}}
\newcommand{\mpifr}{Max-Planck-Institut f\"ur Radioastronomie, Auf dem H\"ugel 69, D-53121 Bonn, Germany \label{mpifr}}
\newcommand{\rutgers}{Department of Physics and Astronomy, Rutgers, the State University of New Jersey, 136 Frelinghuysen Road, Piscataway, NJ 08854-8019, USA \label{rutgers}}
\newcommand{\uchic}{Department of Astronomy \& Astrophysics, The University of Chicago, 5640 S Ellis Ave., Chicago, IL 60637, USA \label{uchic}}
\newcommand{\labm}{Aix Marseille Univ., CNRS, CNES, LAM, Marseille, France 
\label{labm}}
\newcommand{\dura}{Institute for Computational Cosmology, Durham University, South Road, Durham, DH1 3LE, UK \label{dura}}
\newcommand{\durb}{Centre for Extragalactic Astronomy, Durham University, South Road, Durham, DH1 3LE, UK \label{durb}}
\newcommand{\tw}{ThoughtWorks Technologies India Private Limited, Yerawada, Pune 411 006, India \label{tw}}
\newcommand{\ukzn}{Astrophysics Research Centre and School of Mathematics, Statistics \& Computer Science, University of KwaZulu-Natal, Durban 4041, South Africa \label{ukzn}}
\newcommand{\ucarol}{Department of Physics and Astronomy, University of South Carolina, Columbia, SC 29208, USA \label{ucarol}}
\newcommand{\sms}{School of Mathematics, Statistics \& Computer Science, University of KwaZulu-Natal, Durban 4041, South Africa \label{sms}}
\newcommand{\idia}{The Inter-Univ. Institute for Data Intensive Astronomy (IDIA), Dep. of Astronomy, and Univ. of Cape Town, Private Bag X3, Rondebosch, 7701, South Africa, and Univ. of the Western Cape, Dep. of Physics and Astronomy, Bellville, 7535, South Africa \label{idia}}

%%%%%%%%%%%%%%%%%%%%%%%%%%%%%%%%%%%

   \title{\pks: OH and \hi\ at $z=0.89$ and the first MeerKAT UHF spectrum}
   %\subtitle{PKS1830-211: OH and 21cm}
   \author{
        F. Combes\inst{\ref{lerma}}
            \and
        N. Gupta\inst{\ref{iucaa}}
            \and
        S. Muller\inst{\ref{chalmers}}
            \and
        S. Balashev\inst{\ref{ioffe}} 
            \and
        G. I. G. J\'ozsa\inst{\ref{sarao}, \ref{rhodes}, \ref{argl}}
            \and
        R. Srianand\inst{\ref{iucaa}}
            \and
        E. Momjian\inst{\ref{nrao}}
            \and
        P. Noterdaeme\inst{\ref{iap}}
            \and
        H.-R. Kl\"ockner\inst{\ref{mpifr}}
            \and
        A. J. Baker\inst{\ref{rutgers}}
            \and
        E. Boettcher\inst{\ref{uchic}}
            \and
        A. Bosma\inst{\ref{labm}}
            \and
        H.-W. Chen\inst{\ref{uchic}} 
            \and
        R. Dutta\inst{\ref{dura}, \ref{durb}}
            \and
        P. Jagannathan\inst{\ref{nrao}}
            \and
        J. Jose\inst{\ref{tw}}
            \and
        K. Knowles\inst{\ref{ukzn}}
            \and
        J-.K. Krogager\inst{\ref{iap}}
            \and
        V. P. Kulkarni\inst{\ref{ucarol}}
            \and
        K. Moodley\inst{\ref{ukzn}, \ref{sms}}
            \and
        S. Pandey\inst{\ref{tw}}
            \and    
        P. Petitjean\inst{\ref{iap}}
            \and
        S. Sekhar\inst{\ref{nrao}, \ref{idia}}    
          }

    \institute{\lerma \and \iucaa \and \chalmers \and \ioffe \and \sarao \and \rhodes \and \argl \and \nrao \and \iap \and \mpifr \and \rutgers \and \uchic \and \labm \and \dura \and \durb \and \tw \and \ukzn \and \ucarol \and \sms \and \idia}

   \date{Received: December 2020; accepted: February 2021}

% \abstract{}{}{}{}{} 
% 5 {} if structured abstract is chosen
 
  \abstract {
   The Large Survey Project (LSP) "MeerKAT Absorption Line Survey" (MALS) is a blind  \hi\ 21-cm and  OH  18-cm absorption  line  survey in the L- and UHF-bands, with the primary goal to better determine the occurrence of atomic and molecular gas in the circum-galactic and inter-galactic medium, and its redshift evolution. 
   Here we present the first results using the UHF-band, obtained towards the strongly lensed radio source \pks, detecting absorption produced by the lensing galaxy.
   With merely 90\,mins of data acquired on-source for science verification and  processed using the Automated Radio Telescope Imaging Pipeline (ARTIP), we detect in absorption the known \hi\ 21-cm and OH 18-cm main lines at $z=0.89$ at an unprecedented signal-to-noise ratio (4000 in the continuum, in each 6~\kms\, wide channel). 
   For the first time we report the detection at $z=0.89$ of OH satellite lines, so far not detected at $z > 0.25$. We decompose the OH lines into a thermal  and a stimulated contribution, where the 1612 and 1720\,MHz lines are conjugate.
   The total OH 1720\,MHz emission line luminosity is 6100\,L$_\odot$. This is the most luminous known 1720\,MHz maser line. It is also among the highest luminosities for the OH-main lines megamasers. %\citep{darling2002}.
   The absorption components of the different images of the background source sample different light paths in the lensing galaxy, and their weights in the total absorption spectrum are expected to vary in time,  on daily and monthly time scales.
   We compare our normalized spectra with those obtained more than 20 yrs ago, and find no variation, in spite of the high signal-to-noise ratios.
  We interpret the absorption spectra with the help of a lens galaxy model, derived from an N-body hydro-dynamical simulation, with a morphology similar to its optical HST image. The resulting absorption lines depend mainly on the background continuum, and the radial distribution of the gas surface density, for each atomic /molecular species. We show that it is possible to reproduce the observations assuming a realistic spiral galaxy disk, without invoking any central gas outflows. There are, however, distinct and faint high-velocity features in the ALMA millimeter absorption spectra,  that most likely originate from high-velocity clouds or tidal features. These clouds may contribute to broaden the \hi\ and OH spectra.
}
   

\maketitle
%
%-------------------------------------------------------------------
%%%%%%%%%%%%%%%%%%%%%%%%%%%%%
\section{Introduction}
\label{sec:intro}
%%%%%%%%%%%%%%%%%%%%%%%%%%%%%
\pks\ is a highly reddened radio quasar at z=2.51 \citep{lidman1999}, and also the brightest known radio lens in the sky. It was identified as a gravitational lens on the basis of its peculiar radio spectrum and morphology \citep{Rao1988,Jauncey1991}. Its morphology (see the Appendix) reveals two compact radio components, called North-East (NE) and South-West (SW), separated by one arcsecond and surrounded by a low surface-brightness Einstein ring, meaning that the background radio-source and the lens center are well aligned along the line of sight. The two compact components have a relatively flat spectrum, with a spectral index $\sim -0.7$ consistent with synchrotron emission, while the ring spectrum is steeper, such that at low frequency ($<$ 1.7 GHz) the ring is contributing more significantly to the total flux density. Since, in projection the line of sight of \pks\ is very close to the Galactic plane (galactic latitude = $-5.71^\circ$), and confused with stars, for a long time it was difficult to optically identify the lensing galaxy \citep{courbin1998}. The redshift $z=0.89$ of the lensing galaxy was discovered through molecular absorption \citep{wiklind1996}, and it was determined that most of this absorption comes from the SW image, $\sim 3$~kpc from the center \citep{Frye1997}. There is also a much weaker molecular absorption at V=$-$150\,\kms\ towards the NE image, in the opposite direction at 5~kpc from the center of the nearly face-on lensing galaxy \citep{wiklind1998,koopmans2005}. Throughout this paper, the velocity scale is defined with respect to $z=0.88582$ (heliocentric reference frame), which corresponds to the main molecular (e.g., CO, HCO$^+$, and HCN) absorption components detected at mm wavelengths \citep[][]{wiklind1998}.

The apparent size of the lensed images of the quasar at mm wavelengths is a fraction of a milliarcsecond \citep{jin2003,guirado1999}, implying that the molecular absorption could be originating from a single molecular cloud of a few pc in size. However, due to the thickness of the gaseous plane along the line of sight, the pencil beam from the quasar core travels several hundreds of pc in the galaxy, and encounters many clouds along the line of sight, covering a large velocity gradient. This broadens the absorption spectrum because the gas actually involved in the absorption traces the velocity gradient over $\sim$ 1~kpc in the disk. \hi\ 21-cm and OH 18-cm main absorption lines have also been detected at $z=0.89$ \citep{chengalur1999, koopmans2005}. 

It is worth noting that \pks\ also traces an \hi\ 21-cm absorber at $z=0.19$ \citep{lovell1996, allison2017}, albeit with a smaller velocity width. The 90\% of the total 21-cm optical depth ($\Delta {\rm V}_{90}$) of the low redshift absorber is contained within 86\,\kms\ \citep[][]{gupta2020}.

The absorber at $z=0.89$ is particularly rich in molecules and also in dust. A strong absorption at rest-frame 10$\mu$m reveals amorphous or crystalline silicates \citep{aller2012}.  More than sixty molecular species associated with this absorber have been detected towards the SW component, where the inferred H$_2$ column density is $\sim 2 \times 10^{22}$~cm$^{-2}$ \citep{muller2011,muller2014,tercero2020}. In contrast, only 19 species, including atomic hydrogen and carbon, have been detected towards the NE component (at 5~kpc from the center), where the molecular gas has a lower column density by an order of magnitude and is more diffuse. On the contrary, at 18-21~cm wavelengths, it is the NE component which has the stronger absorption signal \citep{chengalur1999,koopmans2005}. The absorption spectrum peaks at $\rm V =-150$~\kms, and is relatively weaker at $\sim 0$\,\kms. This suggests that, towards the center of the lensing galaxy (SW at 3~kpc), the gas phase is mostly molecular and the \hi\ is depleted.
Compared to the mm-absorptions, the broader linewidths of $\sim$ 280 \kms\ and $\sim$ 150 \kms\ (FWHM) for \hi\ and OH cm-absorption, respectively, at $z=0.89$ are explained by the larger thickness of the \hi\ or OH plane, and to some extent by the more extended background radio continuum at lower frequencies.

Such a strong lens system should be useful to determine the Hubble constant, through the cosmography method or time-delay between the images. However, it is then necessary to precisely determine the mass distribution of the lens, which has been difficult, because of extinction and confusion with Milky Way stars \citep{courbin2002,winn2002}. Even the determination of the center of the lensing galaxy is problematic, and two models have been developed, with a bright peak at the center of the lens being a nearby star \citep{courbin2002} or the bulge of the lensing galaxy \citep{winn2002}. 
The detection of several absorption components at different velocities, towards the different images helps to determine the geometry of the lens, which is a nearly face-on spiral galaxy \citep{wiklind1998}. Recently, the detection of a third lensed image of \pks\ with ALMA by \citet{muller2020} has brought more constraints leading to the refinement of the lens model.

%%%%

Due to its very strong radio continuum and absorption lines, \pks\ is also one of the favorite targets for science verification at various radio telescopes. Very recently it was observed as part of the preparation for the MeerKAT Absorption Line Survey (MALS) to demonstrate the spectral line capabilities of MeerKAT. MALS has begun MeerKAT science verification and proper survey observations using the L- and UHF-bands covering 900-1670\,MHz and 580-1015\,MHz, respectively \citep[][]{Gupta17mals}. Owing to the excellent sensitivity of MeerKAT \citep[][]{Jonas16, Camilo18, Mauch20}, the L-band spectrum of \pks\ presented in \citet[][]{gupta2020} provided a robust characterization of \hi\ 21-cm (from $z=0.19$) and OH 18-cm main absorption lines from $z=0.89$. Here, we present the first MeerKAT UHF-band spectrum that simultaneously covers both \hi\ and all four OH 18-cm lines for the $z=0.89$ system. We report for the first time the detection of the OH satellite lines.

%%%%%
This paper is structured as follows. Section \ref{sec:obs} presents the details of the observations and data analysis with ARTIP. The results in terms of optical depths and column densities are quantified in Section \ref{sec:res}, where we discuss the origin of the various absorption components in comparison with the dense molecular gas absorptions, and present a lens-galaxy model. Section \ref{conclu} summarises our conclusions.
%%%%
To compute distances, we adopt a flat $\Lambda$CDM cosmology,
with $\Omega_m$=0.29, $\Omega_\Lambda$=0.71, and the Hubble constant
H$_0$ = 70 \kms Mpc$^{-1}$. At the distance of the \pks\ absorber, an arcsecond corresponds to 7.8\,kpc, in physical units.

%--------------------------------------------------------------------
\section{Observations and data analysis}
\label{sec:obs}

%%%%%%%
The field centered at \pks\ was observed on July 13, 2020 using the MeerKAT-64  array and 32K mode of the SKA Reconfigurable Application Board (SKARAB) correlator. For these UHF-band science verification observations, the total observable bandwidth of 544\,MHz was split into 32768 frequency channels. This corresponds to a frequency resolution of 16.602\,kHz, or 6.1\,\kms\ at the center of the band i.e., 815.9917\,MHz. The correlator dump time was 8\,seconds and the data were acquired for all 4 polarization products, labelled as XX, XY, YX and YY. Of the 64 antennas, 56 participated in these observations. The baseline lengths in the dataset are in the range: 29 - 7300 m. We also observed PKS\,1934-638 and 3C286 for flux density and bandpass calibrations. Since \pks\ is a bonafide gain calibrator for the VLA in the C- and D- array configurations, there was no need to separately observe a complex gain calibrator. The total on-source time on \pks\ and duration of the observations are 90 and 155\,mins, respectively. 
The full dataset in measurement set format is about 3.1\,TB. 

\begin{figure*}
\begin{center}
\includegraphics[trim = {0cm 0.8cm 0.3cm 0.3cm}, width=0.8\textwidth,angle=0]{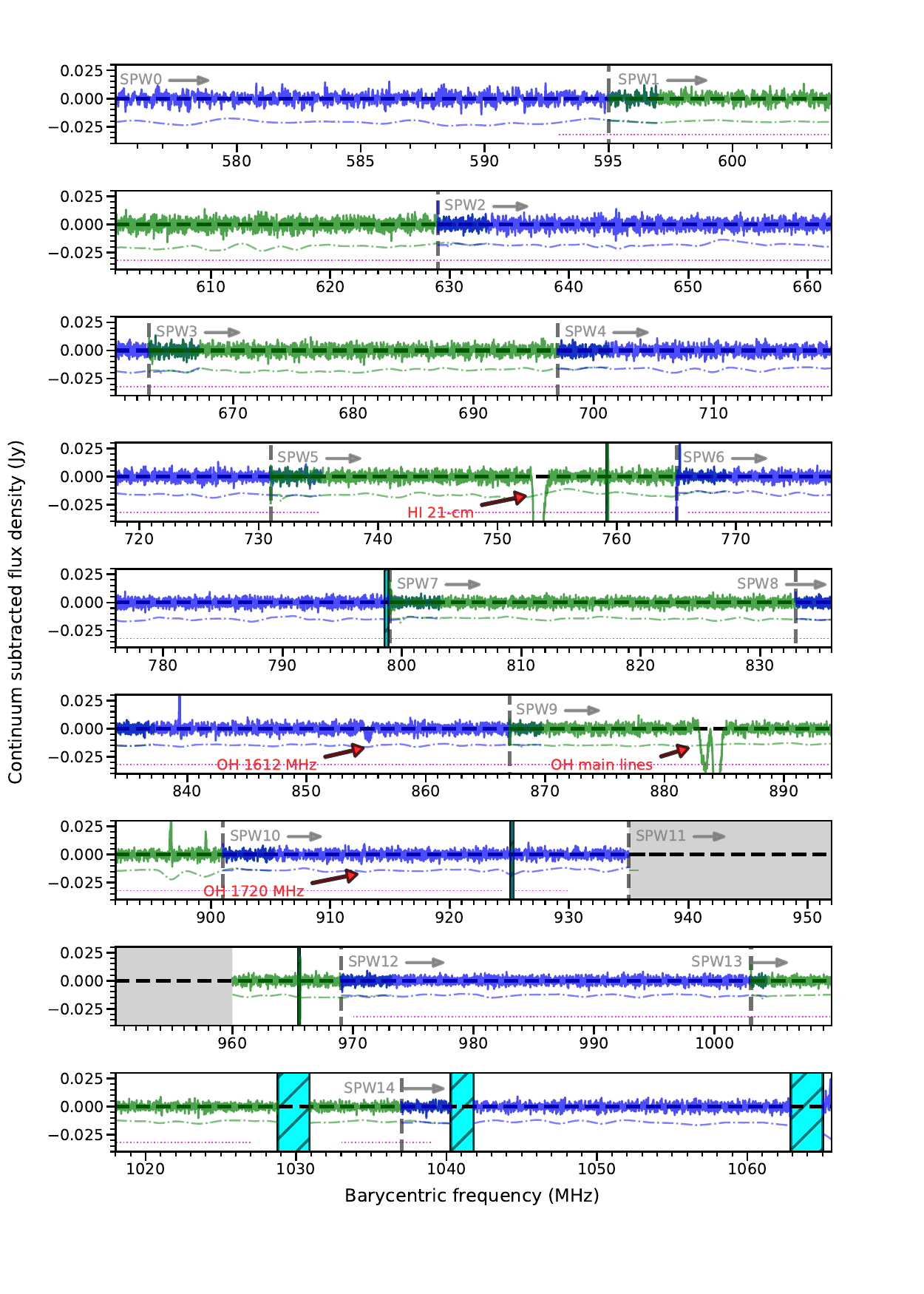}
\vskip+0.0cm
\caption{
	The continuum subtracted Stokes-$I$ spectrum of \pks. The adjacent  Spectral Windows (SPWs) are plotted alternately in blue and green, separated by dashed vertical lines. Gray-shaded regions mark frequency ranges that were masked prior to any calibration. Hatched regions were masked after calibration and imaging to exclude persistent RFI identified during the absorption line analysis.  The dash-dotted blue or green lines represent the error spectrum (5$\times\sigma_{\rm rolling}$). The dotted magenta line marks frequency ranges used for the continuum imaging. The locations of redshifted \hi\ 21-cm and OH 18-cm line frequencies at $z=0.89$ are also indicated.
} 
\label{fig:pks1830full} 
\end{center}
\end{figure*}

Here we are interested only in the Stokes-$I$ properties of the target, therefore for processing we generated a measurement set consisting of only XX and YY polarization products. We also dropped the extreme 1024 frequency channels at both edges of the band. The resulting data set consisting of 30720 channels, was processed on the VROOM cluster at IUCAA using the latest version of the Automated Radio Telescope Imaging Pipeline {\tt ARTIP} based on CASA 5.6.2. The details of ARTIP and data processing steps are provided in \citet[][]{gupta2020}. 
%%%
In short, an initial RFI mask was applied to the data to mask the strongest radio frequency interference (RFI) in the band. Up to this stage, only $\sim$10.0\% of the bandwidth gets flagged.
After this step, the data were further flagged and calibrated using the {\tt ARTIP-CAL} package. The spectral-line processing of calibrated visibilities was done using the {\tt ARTIP-CUBE} package. The final continuum-subtracted spectrum of \pks\ is shown in Fig.~\ref{fig:pks1830full}. The gray-shaded regions in the figure mark the above-mentioned initial RFI mask. For the {\tt ARTIP-CUBE} processing, the frequency band was partitioned into 15 spectral windows (SPW) with an overlap of 256 frequency channels. The unique frequency ranges covered by these measurement sets are marked by dashed vertical lines in Fig.~\ref{fig:pks1830full}. For easier referencing, we refer to these as SPW-0 to -14. 

%%%%%%%%%%%%%%%%%%%%%%%%%%%%%
%%%%%%%%%%%%%%%%%%%%%%%%%%%%%

%%%%%
Recall that prior to any flagging and calibrations the 935 - 960 MHz range was flagged to mask persistent strong RFI. However, the spectral line processing revealed some additional RFI. Specifically, we find the following frequency ranges to be persistently affected by RFI: 759.1 - 759.2, 798.6 - 798.9, 925.1 - 925.3, 965.5 - 965.6, 1028.8 - 1030.9, 1040.3 - 1041.8 and 1062.9 - 1065.0\,MHz (see hatched regions in Fig.~\ref{fig:pks1830full}). We also find a few frequency ranges to be affected by sporadic RFI. These are at 735.275, 765.28, 896.6, 899.294 - 899.404 and 928.0 - 935.0 MHz, and often accompanied by positive / negative spikes in the spectra. The spike at $\sim$839\,MHz is caused by the known \hi\ 21-cm absorption line in the spectrum of the bandpass calibrator \citep[3C286;][]{Wolfe2008}. Nevertheless, as shown in Fig.~\ref{fig:pks1830full} the UHF spectrum is spectacularly clean, and we detect the known \hi\ 21-cm and OH 18-cm main lines at $z=0.89$. Due to the excellent sensitivity, the MeerKAT spectrum has also led to the detection of OH satellite lines (further details in Section~\ref{sec:res}).

%%%%%%
For continuum imaging through the {\tt ARTIP-CONT} package, a more stringent RFI mask to completely exclude band edges and RFI-afflicted regions was applied to the calibrated visibilities (see horizontal dotted lines in Fig.~\ref{fig:pks1830full}). The data were then averaged in frequency over 32 channels ($\sim$0.531\,MHz) and regridded along the frequency axis into 16 distinct spectral windows. We created a widefield broad band 6k$\times$6k continuum image with a pixel size of $3^{\prime\prime}$, spanning $\sim5^\circ$ using {\tt tclean} in 
CASA. The {\tt w-projection} algorithm was used as the gridding algorithm in combination with {\tt Multi-scale Multi-term Multi-frequency synthesis} ({\tt MTMFS}) for deconvolution, with nterms = 2 and four pixel scales to model the extended emission \citep[cf.][]{rau2011, bhatnagar2013, jagannathan2017}. Two rounds of phase-only self-calibration were carried out along with a final round of amplitude and phase self-calibration. Imaging masks were appropriately adjusted using {\tt PyBDSF} between major cycles during imaging and self-calibration \citep[][]{mohan2015}.

%%%%%%%
The final continuum image of \pks\ constructed using {\tt robust=0} weighting has a synthesized beam of $17.4^{\prime\prime}\times13.1^{\prime\prime}$ with a position angle = $+69.0^\circ$. The rms noise in the continuum image is 80\,$\mu$Jy/beam close to the bright radio source at the center and 30\,$\mu$Jy/beam (dynamic range $\sim$380000) away from it.  
The total continuum flux density of the quasar is 11.40 $\pm$ 0.01\,Jy at the reference frequency of 832\,MHz. The quoted uncertainty on the flux density corresponds to errors from the single Gaussian component fitted to the continuum image. Note that the flux density accuracy at these low frequencies is expected to be about $\sim$5\%.  The flux density at 832 MHz is within 1.3\% of the flux density of 11.25\,Jy measured at 1270\,MHz by \citet[][]{gupta2020} from the MeerKAT L-band data acquired on December 19, 2019. However, note that in general the quasar is known to be variable at radio wavelengths\footnote{Over 1996-2016, the flux density at $\sim1.4 - 0.8$\,GHz varied between 10-14\,Jy \citep[cf. Fig.~7 of][]{allison2017}}. The in-band integrated spectral index is $\alpha$ = 0.004 $\pm 0.001$.
%%%%

%%%%%%%%%%%%%%%%%%%%%%%%%%%%%%%%%%%%%%%%%%%%%%%%%%
\section{Results}
\label{sec:res}
%%%%%%%%%%%%%%%%%%%%%%%%%%%%%%%%%%

%%%%%%%%%%%%%%%%%%%%%%%%%%%%%%%
\subsection{OH absorption}
%%%%%%%%%%%%%%%%%%%%%%%%%%%%%%%

%%%%%%%%%%%%%%%%%%
\begin{figure*}
\centering
\includegraphics[trim = {0.5cm 4.5cm 0.5cm 1.0cm}, width=0.80\textwidth,angle=0]{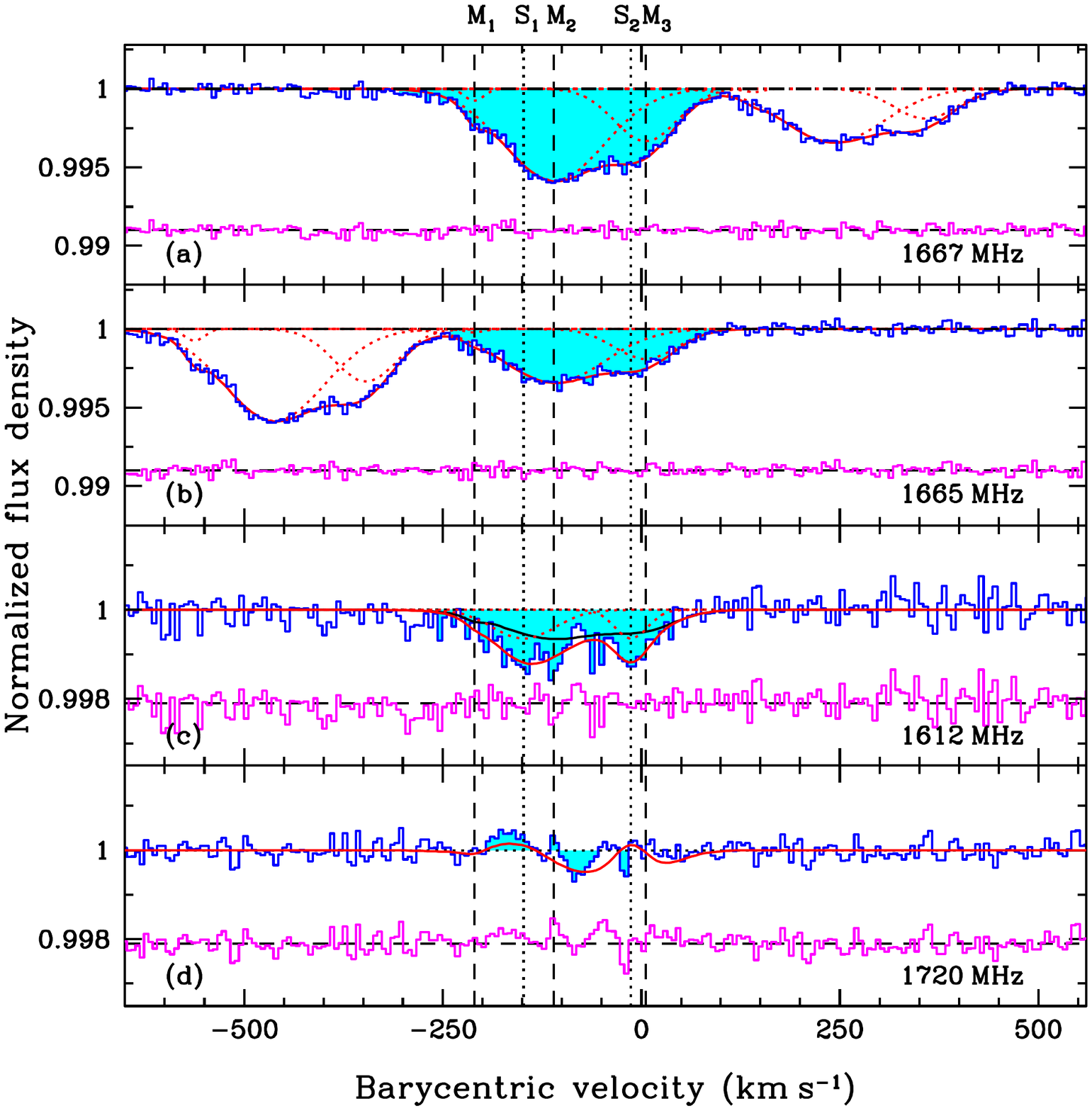}
\vskip+0.0cm
\caption{
MeerKAT Stokes-$I$ spectrum (blue) of OH 18-cm lines at $z=0.89$ towards \pks. The  velocity scale is defined with respect to $z$=0.88582.  The spectral rms in the region close to the 1667\,MHz line is 2.9\,mJy\,beam$^{-1}$\, per 6.1\,\kms\,channel. The vertical dashed and dotted lines (black) represent the locations of the individual Gaussian components fitted to the absorption in main lines (M$_1$, M$_2$ and M$_3$; dotted red lines in (a) and (b)) and the excess absorption in the 1612\,MHz satellite line (S$_1$ and S$_2$; dotted lines in (c); also see Table~\ref{tab:ohfit}), respectively.
The black solid line in panel (c) represents the absorption profile predicted under LTE due to M$_1$, M$_2$ and M$_3$. The difference between the red and black solid lines represent the maser contribution due to pumping, and is conjugate with the 1720\,MHz line in (d): at V=-170\,\kms, the 1720\,MHz line is in emission and 1612\,MHz in absorption, while at V=-80\,\kms, the 1720\,MHz is in absorption, and the 1612\,MHz in emission.
%%%
In all the panels, the total fit and the residuals, arbitrarily offset for clarity, are plotted as solid lines in red and magenta, respectively.
%%%
}
\label{fig:pksohall}
\end{figure*} 
%%%%%%%%%%%%%%%%%%%%%%%%%%%%%

%%%%%%
 The OH main lines with rest frequencies of 1665 and 1667\,MHz are clearly seen in the first MeerKAT UHF-band spectrum (580-1015\,MHz) presented here (see Fig.~\ref{fig:pks1830full}).
We zoom in on these in the top two panels of Fig.~\ref{fig:pksohall}.
 These lines were also detected in the overlapping L-band (900-1670\,MHz) spectrum obtained with MeerKAT on December 19, 2019 \citep{gupta2020}. The absorption obtained from the first UHF spectrum presented here is in good agreement with the L-band profile (see Section~\ref{sec:var} for details on variability).  In our L-band observations, the 1720\,MHz line redshifted to 912\,MHz was also covered but only tentatively detected due to low-SNR and hence not claimed. The  redshifted 1612\,MHz line was outside the L-band. With the present UHF-spectrum, we have clearly detected the 1612\,MHz line (redshifted to 855\,MHz).  Combining the L- and the UHF-band spectra with appropriate rms weights, we now also detect the 1720\,MHz line at a  greater significance. Recall that the redshifted satellite line frequencies are unaffected by RFI (see Section~\ref{sec:obs}).  The profiles of the two satellite lines are presented in the bottom two panels of the figure.

%%%%%%%
This is the first time the OH satellite lines have been detected towards \pks. This is also the highest redshift so far at which the  OH satellite lines have been detected. The detection of all four OH 18-cm lines from the same spectrum allows us to investigate the nature of the absorbing gas without worrying about the line variability. Further, due to the proximity of the 18-cm lines in the frequency space the structure of the background radio continuum illuminating the absorbing gas can be assumed to be the same.

%%%%%%
The integrated optical depths of the main i.e., 1665 and 1667\,MHz lines are 0.729\,$\pm$\,0.012 and 1.301\,$\pm$\,0.013\,\kms, respectively.
 90\% of the total optical depth ($\Delta V_{90}$) of the 1667\,MHz line is contained within 265\,\kms.
%%%%%
For an optically thin cloud, the integrated OH optical depth of the 1667\,MHz line is related to the OH column density $N$(OH) through
%%%%%%
\begin{equation}
N{({\rm OH})}=2.24\times10^{14}~{T_{\rm ex}\over f_{\rm c}^{\rm OH}}\int~\tau_{1667}(v)~{\rm d}v~{\rm cm^{-2}}, 
\label{eqoh}
\end{equation}
%%%%%
where $T_{\rm ex}$ is the excitation temperature in Kelvin, $\tau_{1667}$($v$) is the optical depth of the 1667\,MHz line at 
velocity $v$, and $f_c^{\rm OH}$ is the covering factor \citep[e.g.,][]{Liszt96}.
%%%%
Adopting $T_{\rm ex}$ = 5.14\,K i.e., coupled to the cosmic microwave background (CMB), $T_{\rm CMB}$ at $z=0.89$,
we estimate $N$(OH) = ($1.49 \pm 0.02) \times 10^{15} (T_{\rm ex}/5.14\,{\rm K})(1.0/f_c^{\rm OH}$)\,\cmsq.  Note that this is a surface average, since it is likely that the filling factor is below 1.

%%%%%%%%%%%%%%%%%%%%
\begin{table}
\caption{Multiple Gaussian fits to the OH lines at $z=0.88582$.}
\vspace{-0.4cm}
\begin{center}
\begin{tabular}{cccc}
\hline
\hline
{\large \strut}     Id.    &    Centre     &  $\sigma$            &   $\tau_p$       \\
                           &   (\kms)      &   (\kms)             &   ($10^{-3}$)    \\
\hline
	\multicolumn{4}{c}{1667\,MHz absorption} \\
\hline
                M$_1$      &  -211 $\pm$ 3       & 12 $\pm$ 4       &    0.8 $\pm$ 0.1       \\
                M$_2$      &    -110 $\pm$ 3       &  63 $\pm$ 3       &    5.9 $\pm$ 0.1        \\
                M$_3$      &   6 $\pm$ 3       &  40 $\pm$ 2       &   3.3 $\pm$ 0.3        \\
%%%
\hline
	\multicolumn{4}{c}{1665\,MHz absorption\tablefootmark{a}} \\
\hline
                M$_1$      &  -               &  -               &    0.2 $\pm$ 0.1       \\
                M$_2$      &  -               & -                &    3.4 $\pm$ 0.1       \\
                M$_3$      &  -               & -               &    1.9 $\pm$ 0.2       \\
\hline
\hline
	\multicolumn{4}{c}{Excess 1612\,MHz absorption } \\
\hline
                S$_1$      &  -148 $\pm$ 6             & 39 $\pm$ 6              &    0.6 $\pm$ 0.1       \\
                S$_2$      &  -13 $\pm$ 5             & 21 $\pm$ 5              &    0.6 $\pm$ 0.1       \\
\hline
%%%
\end{tabular}
%%%
\tablefoot{ 
\tablefoottext{a}{The centers and widths of corresponding components fitted to the main lines are tied.}   
	}
\label{tab:ohfit}
\end{center}
\end{table}
%%%%%%%%%%%%%%%%%%%%%%%

%%%%%%
In local thermodynamic equilibrium (LTE), the relative strengths of 18-cm lines are expected to follow the ratio, 1612:1665:1667:1720\,MHz = 1:5:9:1. 
The total integrated optical depths of the main lines are remarkably consistent with the LTE ratio. To explore this further, we model the two main lines using multiple Gaussian components. Assuming that these lines originate from the same gas, we tie the centers and widths of the corresponding components. The overall profiles are reasonably modeled by a three-component fit (reduced $\chi^2\sim1.1$), which is summarized in Table~\ref{tab:ohfit}. The individual components marked as M$_1$, M$_2$ and M$_3$, and the resultant fit along with the residuals are plotted in Fig.~\ref{fig:pksohall} (see panels {(a)} and {(b)}). 
Unsurprisingly, the total absorption in components M$_2$ and M$_3$ which account for $\sim$98\% of the total optical depth is consistent with the LTE ratio. The component M$_1$ is weak and it is difficult to probe its excitation character.
Note that the excitation temperature of the 1.6~GHz OH transitions could be quite different from $T_{\rm CMB}$, especially in the case of far-infrared and/or collisional pumping (see below), and our assumption is conservative, although supported by the LTE ratios of the main lines.

%%%%%%%
If the OH ground state levels are indeed thermalized, the gas producing 1667\,MHz absorption will have 9 times weaker opacity at the 1612\,MHz line velocities. The absorption profile for the 1612\,MHz line based on M$_1$ to M$_3$ is represented by solid (black) lines in Fig.~\ref{fig:pksohall} (panel {(c)}). Clearly, the gas that produces absorption in the main lines can only account for 61\% of the absorption in the satellite line. The remaining excess absorption can be modelled using two components S$_1$ and S$_2$ given in Table~\ref{tab:ohfit}. It appears that the satellite lines have a large contribution due to pumping that leads to non-thermal and possibly even inverted level populations.
%due to pumping that is non-thermal, and even inverted, level population. 
Comparing the two satellite lines, it is clear also that the non-thermal part of the 1612\,MHz line has a deficiency of absorption near -80\,\kms, due to emission (see bottom of Fig.~\ref{fig:pksohall}).

%%%%%%%
Under certain excitation conditions, in particular pumping from far-infrared radiation around 119$\mu$m, the OH levels of the ground state (i.e. $^2\Pi_{3/2}$) can be inverted. In particular the upper level of the  1720\,MHz transition can be highly populated (as the 1612\,MHz transition's lower level), triggering maser emission. Then
the satellite OH lines can exhibit "conjugate" behaviour i.e., as mentioned above the gas exhibiting absorption in one line produces emission 
%in the other, or vice-versa; this peculiar feature 
in the other. This peculiar feature
of the OH levels due to lambda-doubling and hyperfine structure is a precious tool for probing fundamental physical constants \citep{darling2003,chengalur2003,kanekar2018}. Despite the limited SNR,
the 1720\,MHz profile presented in the bottom panel of Fig.~\ref{fig:pksohall} shows both emission and absorption features suggesting such a behavior. 
When the two satellite lines are perfectly conjugate, the sum of their optical depths is zero. The solid red line in the bottom panel of Fig.~\ref{fig:pksohall} corresponds to the sum of {\it (i)} absorption due to M$_1$, M$_2$ and M$_3$ under LTE, and {\it (ii)} absorption due to $S_1$ and $S_2$ assuming perfect conjugate behavior. This simple model provides a reasonable representation of the observed 1720\,MHz profile.

%%%%%
Note that the redshifted 1720\,MHz line frequency is covered in both the L- and UHF-bands of MeerKAT.
Indeed, the spectral features seen in panel {(d)} of Fig.~\ref{fig:pksohall} are also present in our L-band dataset \citep[][]{gupta2020}. The 1720\,MHz profile presented here is already the weighted average of the spectra from the L-band data described in \cite{gupta2020} and the present UHF-band data.
Further investigation into the nature of the satellite lines and the perfection of the conjugate behavior will require a more sensitive spectrum.

%%%%
However, already the 1720\,MHz line profile suggests that the conjugate behavior flips across 
at approximately -100\,\kms, i.e., at velocities $<$-100\,\kms\ 
%it shows in emission and at higher velocities in absorption.
it shows emission and at higher velocities absorption.
%%%
Such flipping has been observed in both galactic and extragalactic systems  \citep[see e.g., cases of Centaurus A and NGC\,253; ][]{langevelde1995, frayer1998}. 
The flipping crucially depends on whether the OH molecules are pumped by the rotational intra-ladder transition at 119\,$\mu$m, in which case 
the 1720\,MHz line shows emission and 1612\,MHz absorption, or by
%the 1720\,MHz line shows in emission and 1612\,MHz in absorption, or by 
the cross-ladder rotational transition at 79\,$\mu$m, in which case the opposite happens \citep[see][especially Fig.~1]{Elitzur1976}.
Since, as explained in \citet[][]{frayer1998}, the intra-ladder transition becomes optically thick much sooner ($N$(OH)/$\Delta$V $\approx 10^{14}$\,\cmsq\,km$^{-1}$\,s) than the cross-ladder transition ($N$(OH)/$\Delta$V $\approx 10^{15}$\,\cmsq\,km$^{-1}$\,s), the observed behavior for \pks\ implies that the gas exhibiting conjugate behavior at velocities $> -100$\,\kms\ has higher OH column density than at $< -100$\,\kms.
Notably, we do not see any significant masing features in 1720\,MHz line at velocities $> -50$\,\kms\ i.e., closer to the center of the lensing galaxy where denser gas tracers such as CO(5-4) or HCO$^+$(2-1) lines towards the SW image have been detected  \citep{muller2014}. 

%%%%%
Finally, the emission component at $-170$\,\kms, which is barely detected at 3$\sigma$ in the 1720\,MHz line, has a total line flux density of 0.037$\pm$0.011\,Jy\,\kms. At $z=0.89$, this corresponds to an enormous line luminosity of $\sim$6100\,L$_\odot$, which is $\sim$20 times more luminous than the previously brightest known 1720\,MHz maser associated with PKS\,1413+135 at $z=0.247$ \citep[][]{kanekar2004,darling2004}. In the latter source, the OH-main lines are detected in absorption, but not at the same velocity as the satellite lines, therefore originating from a different gas component. It is interesting to note that the OH-main lines absorption was not detected at the same time either. \citet{kanekar2004} observed the satellite lines in June 2003 with the WSRT (Westerbork Synthesis Radio Telescope), and the main lines in October 2001 with the GMRT (Giant Metrewave Radio Telescope). \citet{darling2004} observed all four lines in December 2003 with the NRAO Green Bank telescope, but did not detect the main lines, indicating a strong variability.
%The luminosity distance of PKS1830 is indeed 4.6 times higher than PKS1413.
We note that the for \pks, the 1720\,MHz line luminosity is among the highest luminosities of all OH-main line megamasers \citep{darling2002}.

%%%%%%%%%%%%%%%%%%%%%%%%%%%%%%%
\subsection{\hi\ absorption}
%%%%%%%%%%%%%%%%%%%%%%%%%%%%%%%

%%%%%%%%%%%%%%%%%%%%%%%%%%%%%
\begin{figure}
\includegraphics[trim = {0cm 4.8cm 0.3cm 1.2cm}, width=0.48\textwidth,angle=0]{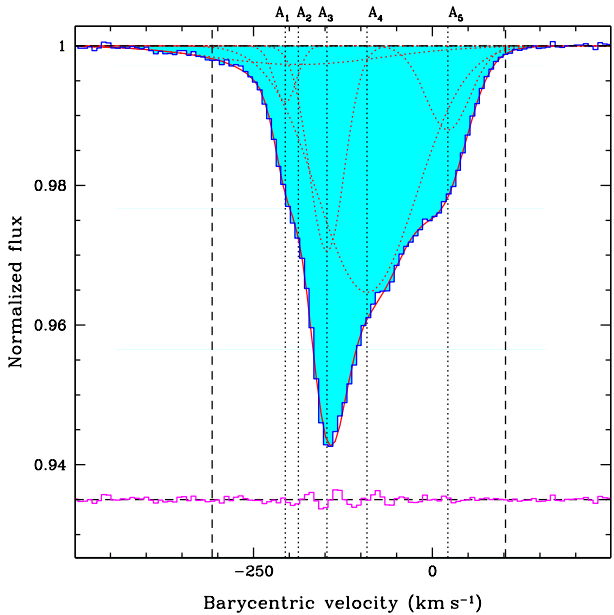}
\vskip+0.0cm  
\caption{
MeerKAT UHF Stokes-$I$ spectrum (blue) of \hi\ 21-cm absorption at $z=0.89$ towards \pks. The spectral rms noise is 3.2\,mJy\,beam$^{-1}$\,channel$^{-1}$. Vertical dotted lines mark positions of Gaussian components, and dashed lines mark the velocity range over which the bulk of OH 1667\,MHz absorption is detected.
} 
\label{fig:pks21cm}
\end{figure} 
%%%%%%%%%%%%%%%%%%%%%%%%%%%%%

%%%%%%%%%%
In Fig.~\ref{fig:pks21cm}, we show \hi\ 21-cm absorption at $z=0.89$. The integrated \hi\ 21-cm optical depth is $\int\tau$dv$ = 10.050\pm 0.019$\,\kms, and the width $\Delta V_{90}$ = 258\,\kms.
For an optically thin cloud the integrated 21-cm optical depth (${\cal{T}} \equiv \int\tau dv$) is related to the neutral hydrogen column density 
$N$(H~{\sc i}), spin temperature $T_{\rm s}$, and covering factor $f_c^{\tiny \hi}$ through,
%%%%%%
\begin{equation}
	N{(\hi)}=1.823\times10^{18}~{T_{\rm s}\over f_{\rm c}^{\tiny \hi}}\int~\tau(v)~{\rm d}v~{\rm cm^{-2}}.
\label{eq21cm}
\end{equation}
%%%%%
Using this we get $N$(\hi) = $(1.83 \pm 0.01)\times10^{21}$ ($T_{\rm s}/100\,{\rm K})(1.0/f_{\rm c}^{\tiny \hi}$)\,\cmsq.

%%%%%%%%%%%%%%%%%%%%
\begin{table}
\caption{Multiple Gaussian fits to \hi\ absorption at $z=0.88582$.}
\vspace{-0.4cm}
\begin{center}
\begin{tabular}{llll}
\hline
\hline

{\large \strut}     Id.    &    Centre     &  $\sigma$            &   $\tau_p$       \\
                           &   (\kms)      &   (\kms)             &   ($10^{-3}$)    \\
\hline
        A$_1$      & -205 $\pm$ 1  &  15 $\pm$ 1    &   8.6  $\pm$ 0.6        \\
        A$_2$      & -187 $\pm$ 40 &  119 $\pm$ 11  &   2.7  $\pm$ 1.1        \\
        A$_3$      & -148 $\pm$ 1  &  24 $\pm$ 1    &   30.0  $\pm$ 0.9       \\
        A$_4$      & -91 $\pm$  2  &  68 $\pm$ 3    &   35.9  $\pm$ 1.4        \\
        A$_5$      &  22 $\pm$ 1    & 31 $\pm$ 1   &   12.2  $\pm$ 0.7       \\

\hline
%%%
\end{tabular}
%%%
%\tablefoot{ 
%\tablefoottext{a}{The centers and widths of main-line corresponding components are tied.}   
%	}
\label{tab:21cmfit}
\end{center}
\end{table}
%%%%%%%%%%%%%%%%%%%%%%%

%%%%%
The bulk of \hi\ absorption can be identified with two components: one centered at $\sim$0 and the other at $-150$\,\kms. These are the velocities associated with mm-wave absorption lines towards the SW and NE components, respectively. Indeed, for the earlier reported 21-cm absorption profiles, which are presented in Figs.~\ref{fig:comp-HI} of Section~\ref{sec:var}, a two-component Gaussian decomposition provided a reasonable fit.
However, the high signal-to-noise ratio of the  MeerKAT spectrum requires at least five Gaussian components (A$_1$ to A$_5$; reduced $\chi^2\sim$1.3) as presented in Table~\ref{tab:21cmfit} and Fig.~\ref{fig:pks21cm}. The remaining structure in the residuals is at the level of $\tau \leq$ 0.0014. It does not necessarily represent discrete physical structures. Therefore, we do not attempt to improve the fit by adding more components. 

%%%%%%
Further, since the Gaussian components obtained above do not necessarily represent individual physical structures, we can also utilize the velocity components previously used to model the OH lines. But this exercise requires more components because there is an additional spectral feature present in the blue wing of \hi\ 21-cm absorption (see vertical dashed lines in Fig.~\ref{fig:pks21cm}).     
%We can retrieve the same two-main velocity components in the OH-absorption spectrum, cf Fig.\ref{fig:pksoh}. 
A rather remarkable result which can be obtained without any component-wise comparison is that the average [OH/HI] column density ratio $\sim$ $8\times10^{-7}$ is slightly higher than the typical ratio ($\sim10^{-7}$) observed in the Galaxy \citep[][]{Li2018}, and also in extra-galactic absorbers \citep{kanekar2002, Gupta2018oh}. For \pks\, this may be primarily due to the fact that the absorption at $\sim$0\,\kms\ i.e., towards the SW component originates from gas which is primarily molecular. 
In Section \ref{sec:var} we will provide evidence indicating that all these values are in perfect agreement with previous results, obtained more than 20 years ago \citep{chengalur1999,koopmans2005}.

%%%%%%%%%%%%%%%%%%%%%%%%%%%%%%%%%%%%%%%%%%
\subsection{Time variability}
\label{sec:var}
%%%%%%%%%%%%%%%%%%%%%%%%%%%%%%%%%%%%%%%%%

 The continuum flux from the background blazar at $z=2.507$ is significantly varying in time, by factors up to 10 in gamma-rays and 2 at radio wavelengths
\citep[e.g.,][]{lovell1998,marti2013}. These variations are seen in all three lensed components i.e. NE, SW, and ring with time-delays. In X-rays, \cite{oshima2001} show that the NE/SW flux ratio varies by factors as high as 7, invoking possible micro-lensing effects.  While the NE and SW components are images of the blazar core, the ring is mainly due to the jet and a bright knot in the jet \citep[e.g.,][]{jin2003}, and there is an observed jet precession period of one year \citep{nair2005}. The delay between the two compact (NE and SW) radio components has been measured to be 27\,days \citep[e.g.,][]{lovell1998,wiklind2001,barnacka2011}.
\citet{barnacka2015} have observed several gamma-ray flares with Fermi-LAT and derived gamma-ray time delays of 23$\pm$0.5 days and 19.7$\pm$1.2\,days, which might be consistent with the radio time delay; however this has been debated \citep[see a reanalysis by][]{abdo2015}. With VLBA observations at 4\, GHz, during a period of 44 days \cite{garrett1997} found highly variable sub-milliarcsecond radio structures in the cores of both lensed images.
It is therefore expected that the variations of the different continuum components will be reflected in the shape of the \hi\ and OH absorption lines through time variability.

\begin{figure}
\includegraphics[width=0.48\textwidth,angle=0]{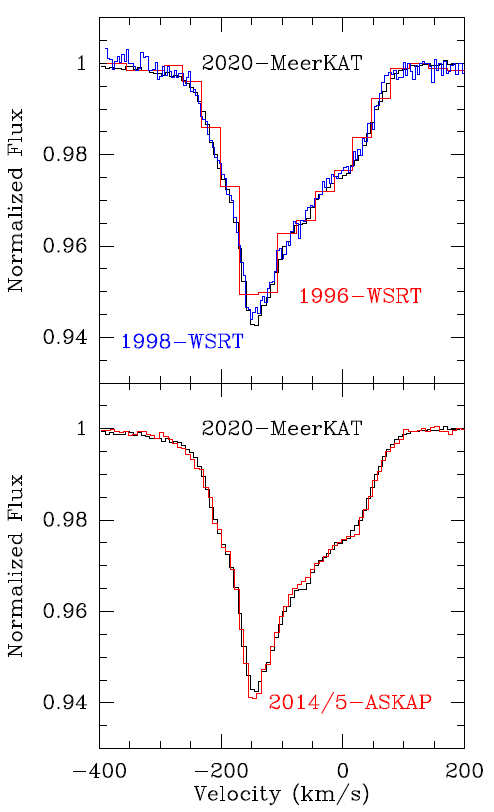}  
\vskip+0.0cm  
\caption{  {\bf Top:}
MeerKAT  \hi\ spectrum (black), taken on July 13, 2020, compared to the \hi\ spectrum (red) observed on November 3, 1996 by \cite{chengalur1999}, and the one (blue) observed on 1998 October 25 by \cite{koopmans2005}, both with the WSRT. 
{\bf Bottom:}
MeerKAT  \hi\ spectrum (black), taken on July 13, 2020, compared to the \hi\ average ASKAP spectrum (red) observed in July 2014 and July-October 2015 by \cite{allison2017}. 
All spectra have been normalized to their observed continuum.
} 
\label{fig:comp-HI}   
\end{figure}

%%%%%%
We have compared all available \hi\ and OH spectra from 1996 with the present MeerKAT ones. In Fig. \ref{fig:comp-HI}, we show the comparison between all \hi\ spectra. It is striking to see that they are all compatible, especially the last ones with high signal-to-noise ratio. No time variation is visible in the line, while the continuum varied by 40\%. The MeerKAT spectrum has a 5-6 times higher signal-to-noise ratio than the old WSRT data by \cite{koopmans2005}. The slight differences can therefore be completely attributed to the higher noise in the latter. The SNR in the WSRT spectrum is 35 at the peak of the absorption, in channels of 7.9~\kms. The absorption peaks at 5\% of the continuum, and the SNR on the continuum is 700. The average ASKAP spectrum has an SNR$\sim$1000 for the continuum, and 50 for the line in channels of 7.3~\kms, and this is the limiting factor in the comparison with our MeerKAT spectrum.  There is a slight difference around V=-60~\kms and V=-150\kms, of max amplitude of 1.5$\times$10$^{-3}$ normalised to the continuum in one channel, and a smaller difference in the two neighboring channels.  But after averaging this leads to a difference significant only at the 1.7$\sigma$ level. Similarly, the OH spectra also show no variation. The comparison between the MeerKAT OH-main line spectra from December 19, 2019 \citep[][]{gupta2020} and July 13, 2020 (this paper) sets a limit of $\Delta \tau_{3\sigma} < 0.0008$. 

%%%%%%%
This lack of variation at cm-wavelength contrasts with the variations detected in the mm-wave absorption spectra  \citep{muller2008,muller2014}. 
We can interpret the above-mentioned constancy at cm-wavelengths with several arguments: first, given the wider extent of the continuum emission at cm-wavelengths, especially in the Einstein ring, the absorbing regions are much larger, whereas the continuum at mm-wavelength consists mainly of two point sources; second, the absorbing medium in the \hi\ component is more diffuse and less fragmented in the dense clouds \citep[][]{Srianand13dib, Gupta18j1243}, such that the motion of possible plasmons (knots) in the background jet will not produce large variations in the absorption signal.
Another factor is the height of the gas plane, which is likely much smaller in the dense molecular phase than in the atomic phase. The \hi\ plane in spiral galaxies is quite thick, up to 1~kpc or more in the flaring outer parts \citep{burton1976,olling1996}. This means that the region illuminated by a point source will range up to $\sim$ 1~kpc, crossing the inclined plane, as discussed
in Sec. \ref{sec:kinmod}. Therefore, the \hi\ absorption spectrum gets blurred over a size of $0.1^{\prime\prime}$, the beam size of some cm-continuum emission studies \citep{patnaik1993,muller2020}(see also Fig.~\ref{fig:pks-cont}). If the gas surface filling factor is relatively high, all sub-clumps over these regions are averaged out, washing out any spatial variation of the order of 0.2 mas.
In addition, due to opacity effects, the cm and mm continuum radiation does not originate from the same volumes: the mm observations probe deeper inside the jet, and hence probably more active regions of the background source.

%%%%%%%%%%%%%%%%%%%%%%%%%%%%%
%%%%%%%%%%%%%%%%%%%%%%%%%%%%%

\subsection{Kinematic model of the spectra}
\label{sec:kinmod}

\cite{nair1993} proposed a detailed model for the lensing galaxy: they suggest it to be a spiral galaxy centered 0.3\arcsec NE of the SW image, of mass 10$^{11}$ M$_\odot$ (rotational velocity $\sim$ 260\,\kms), of low eccentricity i.e., nearly face-on, and with a position angle $\sim$ 12$^\circ$. The discovery of molecular absorption confirmed that it should be gas rich, and the velocity difference between NE and SW images (150\,\kms) constrained its geometry \citep{wiklind1998}. \cite{koopmans2005} made a kinematic model, taking into account the redshift of the source \citep{lidman1999} and more information about the lens image from \cite{courbin2002} and \cite{winn2002}. They concluded that the lens is inclined by i=17-32$^\circ$ on the sky, with a PA between $-15$ and 34$^\circ$. The model also depends on the velocity width of each absorbing component. Strikingly, the absorption in front of the SW image is rather broad, even relatively in the molecular gas. This could originate from non-circular motions since this region of the lensing galaxy is 
closer to the center (much closer than the NE image, where the line width is smaller). Even if the background core of the blazar is a point source in the mm regime, the region involved in the lensing galaxy has an extension of h$\times$ tan(i), where h is the thickness of the plane, and i the inclination of the galaxy. The broadening of the line can therefore be caused by the velocity gradient in this zone.

One of the main uncertainties in the models of \cite{koopmans2005} is the position of the center of the lens. Now that a third image, which was postulated by \cite{nair1993} in the cm-wave radio images, has been detected with ALMA by \cite{muller2020}, the position of the lens is a fixed parameter (very close to the third image position). Thus, it appears that the bright spot towards the center is indeed a star and not the bulge of the lens galaxy, as proposed by \cite{courbin2002}. 
 \cite{koopmans2005} found six best fit models, as a function of six positions of the lens center. Now we know that only one of these is close to reality, the one where their coordinates of the lens center with respect to the NE image (or the distance AC in Fig. \ref{fig:galaxy-model}) are (0.5\arcsec W, 0.45\arcsec S). Their result was then an inclination of i=32$^\circ$ for the lens plane and PA=-15$^\circ$. In the following we fix i=26$^\circ$ and PA=15$^\circ$ to better fit the HST morphology of the lensing galaxy. The difference of inclination may explain why \cite{koopmans2005} selected a velocity dispersion of 39\,\kms for the \hi\ spectrum, while we need 45\,\kms (see below).

%%%%%%%%%%%%%%%%%%%%
\subsubsection{Galaxy model and background continuum}
%
%%%%%%%%%%%%%%%%%%%%%%%%%%%%%
\begin{figure}
\includegraphics[width=0.48\textwidth,angle=0]{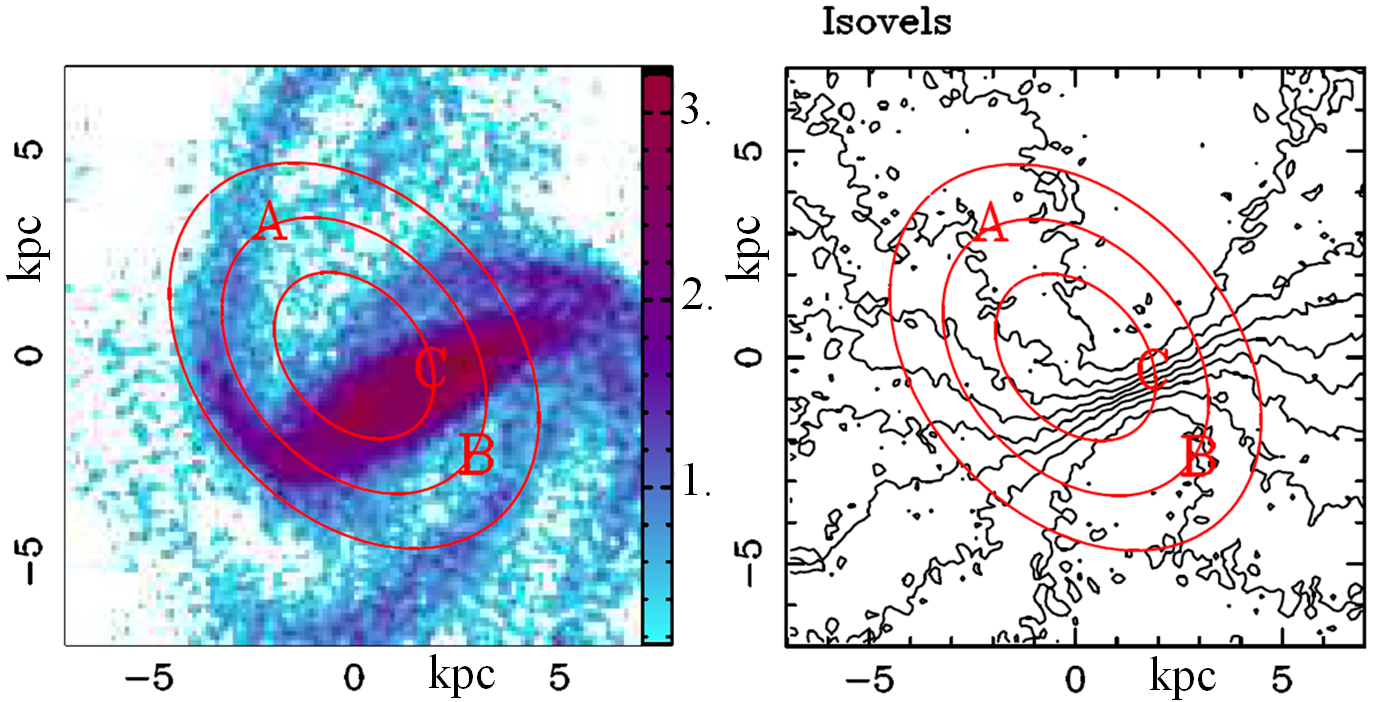}  
\vskip+0.0cm  
\caption{ Galaxy model adopted from an N-body simulation snapshot:
{\bf Left:} the gas surface density, projected on the sky with an inclination of 26$^\circ$ and PA=15$^\circ$,
and {\bf Right:} the corresponding gas isovelocity curves, separated by 24.3 km/s, and ranging from -109.4~\kms to +109.4~\kms. The locations of NE image (A), SW image (B) and 3rd central image (C), close to the lensing-galaxy center, are indicated, together with the schematic contours of the Einstein ring, conspicuous at cm wavelength.
} 
\label{fig:galaxy-model}   
\end{figure} 
%%%%%%%%%%%%%%%%%%%%%%%%%%%%%
%
The HST/WFPC2 F814W image published by \cite{courbin2002} in their Fig. 2 (right) reveals the lensing object as a
barred spiral galaxy, with quite a small bulge. Therefore it is a relatively late-type galaxy. To model the lens, we select a snapshot from an N-body simulation, following the development of a bar and spiral arms, in a late-type object,
described with stellar, gas and dark matter components, appropriately treating star formation and feedback. We chose the snapshot to have a morphology very similar to the HST image. The initial galaxy is a gSb model, described in \cite{combes2008} and \cite{chilingarian2010}. We project the model on the sky with an inclination of 26$^\circ$ and a PA of 15$^\circ$, which are compatible with the above determinations. The resultant gas surface density and the velocity field are plotted in Fig. \ref{fig:galaxy-model}.
The 5\,GHz image obtained using MERLIN (Multi-Element Radio Linked Interferometer Network) is among the best cm-wavelength continuum image of the radio source \citep{patnaik1993}. Although the redshifted \hi\ and OH line frequencies (750-900~MHz) are much lower, we used this high-resolution map with a synthesized beam of 0.1\,arcsec as the base radio continuum model. Fortunately, the spectral index is not very steep between 400 MHz and 5 GHz (see multiple measurements between 4.8 GHz and 408 MHz with the Australia Telescope Compact Array i.e., ATCA and the Parkes telescope,  in the NASA/IPAC Extragalactic Database i.e., NED and Section~\ref{sec:obs}), at least for the main images, although it could be steeper for the Einstein ring. We will also compare the absorption spectra obtained at low frequencies with those in the millimeter domain, where the extent of continuum emission is very different due to the weaker Einstein ring component. The three compact images of the background quasar are also detected with ALMA with a beam of 0.056\,arcsec \citep{muller2020}. The adopted continuum maps for our cm- and mm-wave models are presented in the Appendix (Fig.~\ref{fig:pks-cont}).

%%%%%%%%
\subsubsection{Computations of absorption maps and spectra}
 To reproduce the observations with the above continuum models, we built data cubes by projecting the models
on the sky, with the fixed inclination and position angles, multiplying at each spatial pixel the gas surface density with the background continuum, and thereby 
computing the line-of-sight velocity distribution. We select the same spatial pixel sizes as
the continuum maps i.e., 0.028\,arcsec for the cm-wave, and 0.014\,arcsec for the mm one (see Fig.~\ref{fig:pks-cont}). The adopted spectral channels are of width 6.25\,\kms, for both.

%%%
The main parameters to vary are related to the radial distribution of the various atomic and molecular species considered in the absorption maps and the spectra. The \hi\ distribution is well known to be depleted at the center of spiral galaxies. There are
some extreme examples, such as NGC~628, IC~342, M83, M101 or NGC~7331, with a central
H$_2$/\hi\ surface density ratio of $\sim$100 as can be seen in the THINGS \hi\ \citep[][]{walter2008} and the Heracles
CO surveys \citep{leroy2009}. Some of these radial distributions are computed in \cite{casasola2017}.
We therefore selected an \hi\ radial distribution depleted at the center, such that it gives more absorption weight to the NE(A) image with respect to the SW(B) image. Indeed, the distance of A from the center of the lensing galaxy is 5.3\,kpc, while B is only at 2.4\,kpc from the center
(see also Sec. \ref{sec:intro}). The contribution to the absorption spectrum from B is towards the V=0~\kms\, component while A is contributing to the V=$-$150~\kms\, component. The third image C and the average of the Einstein ring at cm-wavelengths have a continuum contribution of the order of 1\% of the flux density of A or B. Recall that the flux densities of A and B are comparable. 

%%%%%
For the \hi\ component, we selected a power-law radial distribution, as $f=(r/r_0)^\gamma$, with a normalizing scale r$_0$=3.89 kpc, equal to the Einstein ring radius. The gas surface density of the galaxy model was multiplied by this factor f, keeping the underlying bar/spiral structure as shown in Fig. \ref{fig:galaxy-model}. Different radial distributions will lead to different shapes of the total absorption spectrum and in particular different ratios between the two main velocity components. The best fit was obtained with $\gamma=2.5$, which is very similar to the distributions selected by \cite{koopmans2005}. 

%%%%
The velocity widths of \hi\ components are due to two factors: first the intrinsic gas dispersion, and second the averages of all gas rotational velocities in the plane of the lens, seen along the line of sight towards the continuum emission. The strongest continuum sources are the images of the background quasar core, and their angular sizes are quite small ($\sim$mas), in fact smaller than the synthesized beam sizes. However, the line of sight towards these core images traverses the inclined lens galaxy plane ($i=26^\circ$), which has a thickness h. Consequently, regions along h$\times$ tan(i)\,$\sim 0.5\times$\,h are illuminated on the plane, with the corresponding velocity gradient. The width of the spectrum can therefore include velocity gradients over 0.1\,arcsec, which is the resolution of the MERLIN continuum map. Hence, there is no need to deconvolve the continuum map for the model. 
Once the velocity gradients are taken into account, the simulated \hi\ components are still too narrow, and it is necessary to add a velocity dispersion of $\sigma_v= 45$~\kms. This might be due to extra-planar \hi\ gas, frequently observed in nearby spiral galaxies in a few-kpc thick layer, with lagging rotational velocity, and equivalent dispersion up to 30~\kms\,  \citep{marasco2019}.

\begin{figure}
\includegraphics[width=0.48\textwidth,angle=0]{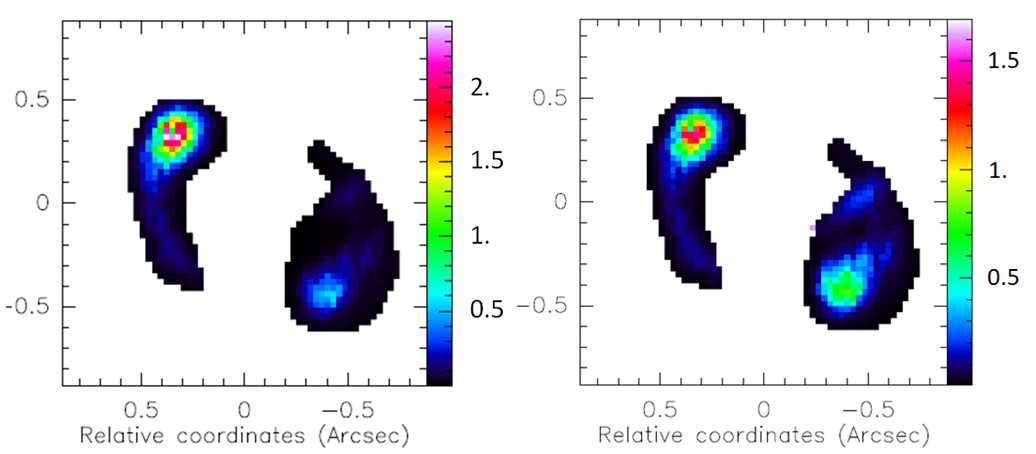}  
\vskip+0.0cm  
\caption{Maps of the product (arbitrary units) between the MERLIN continuum intensity and the optical depth from the galaxy model:
{\bf Left:} for the \hi\ line, where a central depletion was assumed for the best fit
and {\bf Right:} for the OH-1667~MHz line, with a more concentrated distribution. The axes are labelled in arcsec. We note a big difference in the SW component.
} 
\label{fig:pks-maps1}   
\end{figure}

%%%%
We also performed a similar computation for the OH 18-cm main lines, in particular the stronger 1667~MHz line, but with a more centrally-concentrated radial distribution, using a power-law index $\gamma=0.7$. Although the OH does not suffer from the \hi\ depletion towards the center (indeed, there is nothing like the phase change from \hi\ to H$_2$ at high density), the OH radical is still distributed in a thick plane like the HI, a plane which is flaring with radius, and hence increases the velocity coverage of the absorption. This motivates the choice of $\sigma_v= 45$~\kms\ for OH as well and suggests that the OH might also be sharing the extra-planar \hi\ layer. The resulting 2D maps and the spectra are shown in Figs. \ref{fig:pks-maps1} and \ref{fig:pks-lines}. The 2D maps indicate in particular the relative contributions of the various continuum regions in the absorption spectra. 

Most of the absorption is coming from the lines of sight towards the core of the background source. To test the influence of the Einstein ring, whose flux could be relatively higher at low frequency, we have artificially boosted its continuum emission by factors 3-10 with respect to the core images, but there was no significant impact on the main line components.
The various contributions of the different continuum components to the total absorption spectrum are presented in Appendix (\ref{fig:fig-mod}). The ring is contributing a broad component. Although the ring contribution, even at low frequency, should be lower than 10\%, it can broaden the \hi\ and OH spectra, and be responsible in part for their wide wings. Another contribution to the wings may be the high velocity clouds, which will be discussed in Sec. \ref{sec:mm-lines}.

%%%%%%%%%%%%%%%%%%%%%%%%%%%%%
%%%%%%%%%%%%%%%%%%%%%%%%%%%%%

%%%%%%%%%%%%%%%
\subsection{Distribution of the gas absorbing the mm continuum}
\label{sec:mm-lines} 

As mentioned earlier, the continuum emission at millimeter wavelengths is much more compact than at cm, with all emission arising from the compact NE and SW images and little contribution from the ring. It is therefore interesting to compare the absorption spectra between the cm and mm domain, as this can give us constraints on the distribution of gas in the disk. For comparison, we take spectra of several species observed with ALMA, namely HCO$^+$, H$_2$O, CH$^+$, and ArH$^+$ (see below and \citealt{muller2014,muller2015,muller2017}).

\begin{figure}
\includegraphics[width=0.48\textwidth,angle=0]{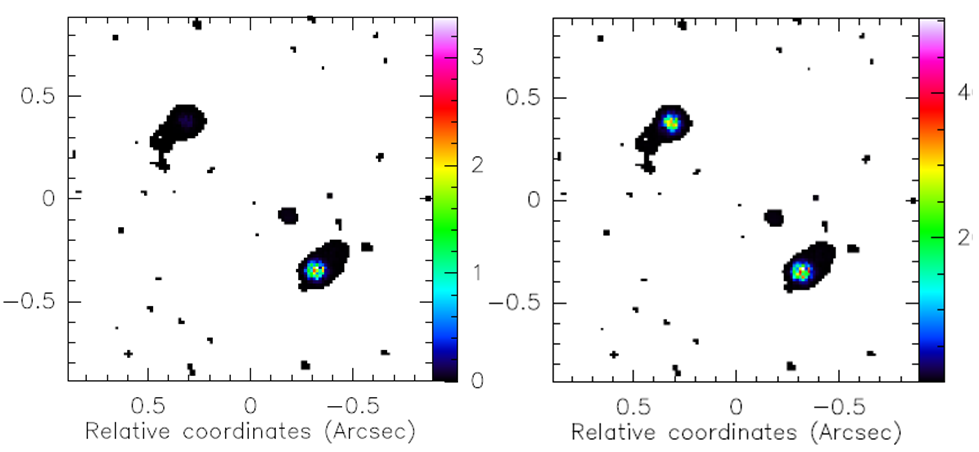}  
\vskip+0.0cm  
\caption{Maps of the product (arbitrary units) between the ALMA continuum intensity and the optical depth from the galaxy model:
{\bf Left:} for the HCO$^+$ line, very similar to the H$_2$O case, both with a highly concentrated distribution
and {\bf Right:} for the ArH$^+$ line, with a more extended distribution (see text). The axes are labelled in arcsec. We note a big difference in the NE component.
} 
\label{fig:pks-maps2}   
\end{figure}

\begin{figure}
\includegraphics[width=0.48\textwidth,angle=0]{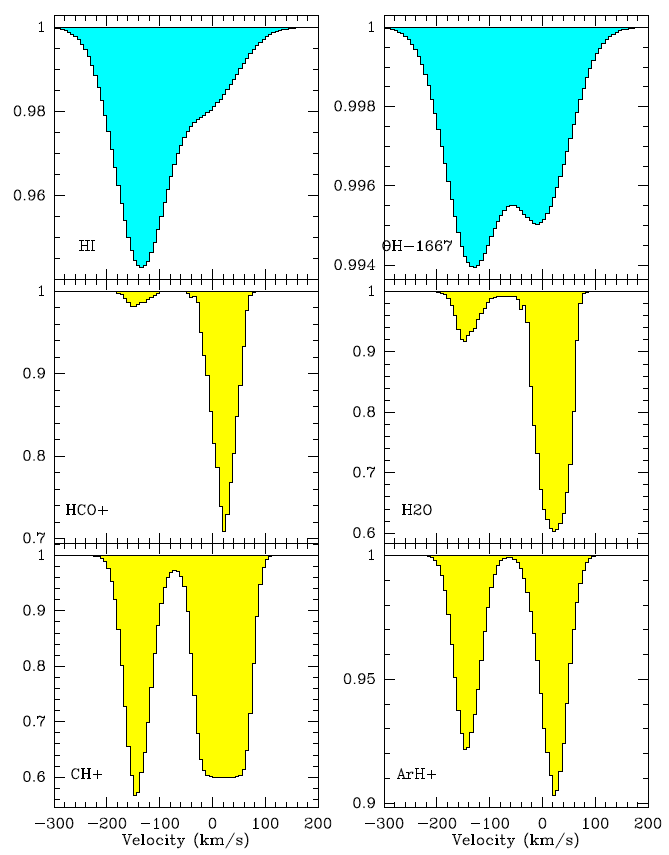}  
\vskip+0.0cm  
\caption{Absorption line profiles derived from the model: the global spectrum has been computed over the whole continuum map,
{\bf Top:} for the \hi\ and OH lines,
{\bf Middle:} the HCO$^+$ and H$_2$O lines
and {\bf Bottom:} for the CH$^+$ and ArH$^+$ ones. The spectra are normalised with the total continuum level (NE+SW+Einstein ring).}
\label{fig:pks-lines}   
\end{figure}

\begin{figure}
\includegraphics[width=0.48\textwidth,angle=0]{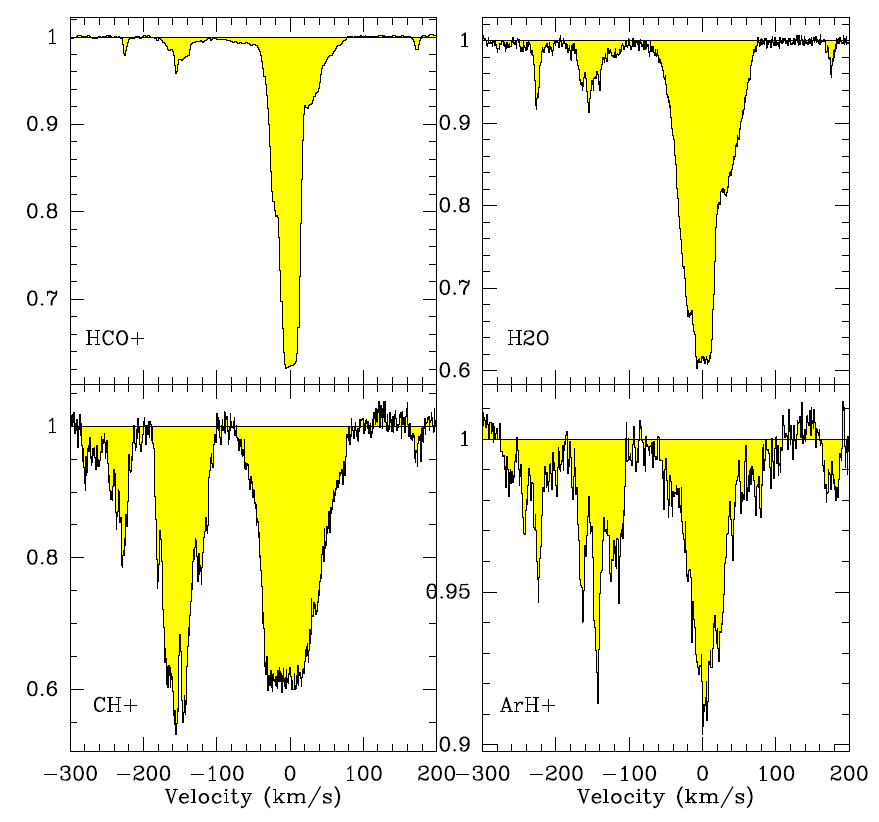}  
\vskip+0.0cm  
\caption{Absorption lines obtained with ALMA towards \pks\,
{\bf Top:} for the HCO$^+$ (2014 Aug 27) and H$_2$O (2016 March 5) species
and {\bf Bottom:} for the CH$^+$ (2015 May 20)  and ArH$^+$ (2015 May 19) ones. The spectra are normalised by the total continuum level (NE+SW+Einstein ring).
We have assumed a flux ratio between NE and SW images of 1.5.}
\label{fig:fig-alma}   
\end{figure}

The widths of the various velocity components are much narrower, and this could be due both to the smaller sizes of the core images, and the smaller thickness of the dense molecular gas plane. We therefore computed the model with twice more spatial resolution in linear size i.e., four times in area, using the ALMA continuum map, which fortunately also has a factor of two better spatial resolution (see Fig\,\ref{fig:pks-cont} in Appendix~\ref{app:contmaps}). To reproduce the observations, the intrinsic dispersion of the gas (6\,\kms) in the N-body hydro simulation is sufficient, and an added convolution is not necessary. We attempt to reproduce absorption from HCO$^+$ which  represents a generically mildly optically-thick species, the H$_2$O molecule which is 
 a highly optically thick one, and ArH$^+$ which should be a tracer of the \hi\, but with a different continuum illumination than at cm-wavelengths. The ArH$^+$ absorption appears optically thin, but there is another species, CH$^+$, tracing a gas with intermediate molecular fraction, which is highly optically thick. The ALMA spectra, combined to a beam encompassing all the continuum emission, are represented in Fig. 
 \ref{fig:fig-alma}. Both HCO$^+$ and H$_2$O have a highly concentrated radial distribution. We therefore selected the exponential profile, with a factor f=exp(-$\gamma$ r/r$_0$), with the same r$_0$=3.89 kpc, as before.
For the HCO$^+$ and H$_2$O distribution, the best fit was obtained with an exponential law, with $\gamma$=2.5. The main difference in their profiles is due to their different optical thickness. In addition, for H$_2$O, it was necessary to deplete the central gas in front of the third image C, to avoid an excessively broad absorption. Indeed, C is close to the galaxy centre, and the velocity gradient at this location reached the maximum of 220 \kms\ in projection. The resulting 2D maps and the spectra are shown in Figs. \ref{fig:pks-maps2} and \ref{fig:pks-lines}.

For the CH$^+$ distribution, the best fit was obtained with an exponential of $\gamma$=1.7, and for ArH$^+$, $\gamma$=0.1. The latter is more extended, with a distribution more similar to the atomic one. In addition, the velocity components are broader, and we selected a convolution width $\sigma_v$= 15~\kms. The latter is three times less than the \hi\ dispersion, but still higher than that of the molecular gas.

The profiles of Fig. \ref{fig:fig-alma} reveal small extra features with velocities lower than $-200$ \kms, or higher than 100 \kms. These velocities do not exist in our galaxy model, and should come either from high-velocity clouds in the halo of the galaxy, or extra tidal arms coming from a possible galaxy interaction. The line of sight of \pks\ falls at low latitude towards the Galactic center (l,b=12.2$^\circ$,-5.7$^\circ$), and the region is too crowded to see tidal tails or a possible companion.
It is possible that the broad \hi\ components include these high-velocity clouds, and are the cause of the high velocity dispersion required by the model.

In summary, there is a wide range of central concentrations for the various species, between HCO$^+$ the most concentrated and \hi\ the most extended tracer. Furthermore there is a large range of optical thickness and of gas layer height, implying a large range of velocity width, that can explain the various morphologies and intensities of spectral features.

%%%%%%%%%%%%%%%%%%%%%%%%%%%%%
%%%%%%%%%%%%%%%%%%%%%%%%%%%%%

%%%%%%%%%%%%%%%%%%%%%%%%%%%%%
%%%%%%%%%%%%%%%%%%%%%%%%%%%%%

%%%%%%%%%%%%%%%%%%%%%%%%%%%%%
%%%%%%%%%%%%%%%%%%%%%%%%%%%%%

\section{Conclusions}
\label{conclu}
   The MALS LSP project on the MeerKAT array has carried out its first science verification observations in the UHF-band, covering the 580 to 1015 MHz frequency range. 
   The brightest known radio lens, \pks\,, has been observed and detected with high signal-to-noise ratio (SNR$\sim$ 4000 in channels of 6 \kms).
   While the \hi\ 21-cm and OH main lines at $z=0.89$ were already known, due to the excellent sensitivity of MeerKAT we have also been successful in detecting the OH 18-cm satellite lines.
   %%%
   We find that while the strength of the OH main lines is consistent with the LTE ratio, the satellite lines show conjugate behavior indicative of radiative pumping due to far-infrared radiation.
   %%%
   The OH emission component in the 1720\,MHz line, although at only 3$\sigma$, has the total line luminosity of 6100\,L$_\odot$, i.e., $\sim$20 times more luminous than the previously brightest known 1720\,MHz maser, associated with PKS\,1413+135 at $z=0.247$ \citep[][]{kanekar2004,darling2004}. A detailed radiative model will be investigated further in the future, when the corresponding lines are re-observed with even higher SNRs.
   
   Contrary to the absorption features detected in several molecules at mm wavelengths, peaking at $\rm V=0$\,\kms\, (i.e. the SW image), the main absorbing component at cm-wavelengths is at a velocity of -150~\kms\, with respect to $z=0.88582$, corresponding to the NE image. We compared our spectrum with those previously obtained with the WSRT \cite[SNR$\sim$ 700 with respect to the continuum and 35 at the peak of the line  in][]{koopmans2005}. The spectra reveal no variation, within an SNR of 35 relative to the depth of the absorption. No variation was seen also with respect to the ASKAP spectrum of \cite{allison2017} with SNR$\sim$1000 with respect to the continuum and 50 at the peak of the line.
   
   We have built a realistic lens galaxy model, from an N-body hydro simulation of an Sb-type galaxy, developing a bar and spiral arms. The galaxy is projected to obtain a similar morphology as the HST image. Combining with the cm and mm continuum images of the background quasar obtained with best spatial resolution, we have computed data cubes to simulate the observed absorption features. The \hi\ and OH absorption spectra are well reproduced with the main velocity field of the galaxy model. An intrinsic velocity dispersion of $\sigma_v$=45\,\kms\ has to be added to account for the broad widths, which can be explained by the presence of an extra-planar \hi\ and OH gas, in addition to a thick and flaring plane. For the HCO$^+$ and H$_2$O absorption lines, only the dispersion of 6\,\kms\ yielded by the simulated dense molecular gas is sufficient to account for the observed profiles. The molecular gas component is much more centrally concentrated, while the \hi-component is depleted in the galaxy center. There is no need for any outflow component coming from the galaxy plane. There exist however some distinct and faint extra features, which can be interpreted as high-velocity clouds. To account for the ArH$^+$ absorption lines, observed with ALMA, a higher velocity dispersion ($\sigma_v$=15\,\kms) and a more extended distribution, similar to the atomic component, has to be selected.  

%%%%%%%%%%%%%%%%%%%%%%%%%%%%%%%%%%%%%%%%%%%%%%%%%%%%%%%%%%%%%%%%
\begin{acknowledgements}
We heartfully thank the SARAO's team of engineers and commissioning scientists for years of intense and successful work towards delivering the wonderful MeerKAT telescope.  We also thank the referee for helpful comments, that improved the clarity of the paper.
The MeerKAT telescope is operated by the South African Radio Astronomy Observatory, which is a facility of the National Research Foundation, an agency of the Department of Science and Innovation. KM acknowledges support from the National Research Foundation of South Africa.
The MeerKAT data were processed using the MALS computing facility at IUCAA (https://mals.iucaa.in/releases).
The Common Astronomy Software Applications (CASA) package is developed by an international consortium of scientists based at the National Radio Astronomical Observatory (NRAO), the European Southern Observatory (ESO), the National Astronomical Observatory of Japan (NAOJ), the Academia Sinica Institute of Astronomy 
and Astrophysics (ASIAA), the CSIRO division for Astronomy and Space Science (CASS), and the Netherlands Institute for Radio Astronomy (ASTRON) under the guidance of NRAO.
The National Radio Astronomy Observatory is a facility of the National Science Foundation operated under cooperative agreement by Associated Universities, Inc. We made use of the
NASA/IPAC Extragalactic Database (NED).

\end{acknowledgements}
%%%%%%%%%%%%%%%%%%%%%%%%%%%%%%%%%%%%%%%%%%%%%%%%%%%%%%%%%%%%%%%%%

\bibliographystyle{aa}
\bibliography{mybib.bib}

\begin{thebibliography}{67}
\expandafter\ifx\csname natexlab\endcsname\relax\def\natexlab#1{#1}\fi

\bibitem[{{Abdo} {et~al.}(2015){Abdo}, {Ackermann}, {Ajello}, {Allafort},
  {Amin}, {Baldini}, {Barbiellini}, {Bastieri}, {Bechtol}, {Bellazzini},
  {Blandford}, {Bonamente}, {Borgland}, {Bregeon}, {Brigida}, {Buehler},
  {Bulmash}, {Buson}, {Caliandro}, {Cameron}, {Caraveo}, {Cavazzuti}, {Cecchi},
  {Charles}, {Cheung}, {Chiang}, {Chiaro}, {Ciprini}, {Claus}, {Cohen-Tanugi},
  {Conrad}, {Corbet}, {Cutini}, {D'Ammando}, {de Angelis}, {de Palma},
  {Dermer}, {Drell}, {Drlica-Wagner}, {Favuzzi}, {Finke}, {Focke}, {Fukazawa},
  {Fusco}, {Gargano}, {Gasparrini}, {Gehrels}, {Giglietto}, {Giordano},
  {Giroletti}, {Glanzman}, {Grenier}, {Grove}, {Guiriec}, {Hadasch},
  {Hayashida}, {Hays}, {Hughes}, {Inoue}, {Jackson}, {Jogler},
  {J{\'o}hannesson}, {Johnson}, {Kamae}, {Kn{\"o}dlseder}, {Kuss}, {Lande},
  {Larsson}, {Latronico}, {Longo}, {Loparco}, {Lott}, {Lovellette}, {Lubrano},
  {Madejski}, {Mazziotta}, {Mehault}, {Michelson}, {Mizuno}, {Monzani},
  {Morselli}, {Moskalenko}, {Murgia}, {Nemmen}, {Nuss}, {Ohno}, {Ohsugi},
  {Paneque}, {Perkins}, {Pesce-Rollins}, {Piron}, {Pivato}, {Porter},
  {Rain{\`o}}, {Rando}, {Razzano}, {Reimer}, {Reimer}, {Reyes}, {Ritz},
  {Romoli}, {Roth}, {Saz Parkinson}, {Sgr{\`o}}, {Siskind}, {Spandre},
  {Spinelli}, {Takahashi}, {Takeuchi}, {Tanaka}, {Thayer}, {Thayer},
  {Thompson}, {Tibaldo}, {Tinivella}, {Torres}, {Tosti}, {Troja}, {Tronconi},
  {Usher}, {Vandenbroucke}, {Vasileiou}, {Vianello}, {Vitale}, {Waite},
  {Werner}, {Winer}, \& {Wood}}]{abdo2015}
{Abdo}, A.~A., {Ackermann}, M., {Ajello}, M., {et~al.} 2015, \apj, 799, 143

\bibitem[{{Aller} {et~al.}(2012){Aller}, {Kulkarni}, {York}, {Vladilo},
  {Welty}, \& {Som}}]{aller2012}
{Aller}, M.~C., {Kulkarni}, V.~P., {York}, D.~G., {et~al.} 2012, \apj, 748, 19

\bibitem[{{Allison} {et~al.}(2017){Allison}, {Moss}, {Macquart}, {Curran},
  {Duchesne}, {Mahony}, {Sadler}, {Whiting}, {Bannister}, {Chippendale},
  {Edwards}, {Harvey-Smith}, {Heywood}, {Indermuehle}, {Lenc}, {Marvil},
  {McConnell}, \& {Sault}}]{allison2017}
{Allison}, J.~R., {Moss}, V.~A., {Macquart}, J.~P., {et~al.} 2017, \mnras, 465,
  4450

\bibitem[{{Barnacka} {et~al.}(2015){Barnacka}, {Geller}, {Dell'Antonio}, \&
  {Benbow}}]{barnacka2015}
{Barnacka}, A., {Geller}, M.~J., {Dell'Antonio}, I.~P., \& {Benbow}, W. 2015,
  \apj, 809, 100

\bibitem[{{Barnacka} {et~al.}(2011){Barnacka}, {Glicenstein}, \&
  {Moudden}}]{barnacka2011}
{Barnacka}, A., {Glicenstein}, J.~F., \& {Moudden}, Y. 2011, \aap, 528, L3

\bibitem[{{Bhatnagar} {et~al.}(2013){Bhatnagar}, {Rau}, \&
  {Golap}}]{bhatnagar2013}
{Bhatnagar}, S., {Rau}, U., \& {Golap}, K. 2013, \apj, 770, 91

\bibitem[{{Burton}(1976)}]{burton1976}
{Burton}, W.~B. 1976, \araa, 14, 275

\bibitem[{{Camilo} {et~al.}(2018){Camilo}, {Scholz}, {Serylak}, {Buchner},
  {Merryfield}, {Kaspi}, {Archibald}, {Bailes}, {Jameson}, {van Straten},
  {Sarkissian}, {Reynolds}, {Johnston}, {Hobbs}, {Abbott}, {Adam}, {Adams},
  {Alberts}, {Andreas}, {Asad}, {Baker}, {Baloyi}, {Bauermeister}, {Baxana},
  {Bennett}, {Bernardi}, {Booisen}, {Booth}, {Botha}, {Boyana}, {Brederode},
  {Burger}, {Cheetham}, {Conradie}, {Conradie}, {Davidson}, {De Bruin}, {de
  Swardt}, {de Villiers}, {de Villiers}, {de Villiers}, {de Villiers}, {De
  Waal}, {Dikgale}, {du Toit}, {du Toit}, {Esterhuyse}, {Fanaroff}, {Fataar},
  {Foley}, {Foster}, {Fourie}, {Gamatham}, {Gatsi}, {Geschke}, {Goedhart},
  {Grobler}, {Gumede}, {Hlakola}, {Hokwana}, {Hoorn}, {Horn}, {Horrell},
  {Hugo}, {Isaacson}, {Jacobs}, {Jansen van Rensburg}, {Jonas}, {Jordaan},
  {Joubert}, {Joubert}, {J{\'o}zsa}, {Julie}, {Julius}, {Kapp}, {Karastergiou},
  {Karels}, {Kariseb}, {Karuppusamy}, {Kasper}, {Knox-Davies}, {Koch},
  {Kotz{\'e}}, {Krebs}, {Kriek}, {Kriel}, {Kusel}, {Lamoor}, {Lehmensiek},
  {Liebenberg}, {Liebenberg}, {Lord}, {Lunsky}, {Mabombo}, {Macdonald},
  {Macfarlane}, {Madisa}, {Mafhungo}, {Magnus}, {Magozore}, {Mahgoub}, {Main},
  {Makhathini}, {Malan}, {Malgas}, {Manley}, {Manzini}, {Marais}, {Marais},
  {Marais}, {Maree}, {Martens}, {Matshawule}, {Matthysen}, {Mauch}, {McNally},
  {Merry}, {Millenaar}, {Mjikelo}, {Mkhabela}, {Mnyand u}, {Moeng}, {Mokone},
  {Monama}, {Montshiwa}, {Moss}, {Mphego}, {New}, {Ngcebetsha}, {Ngoasheng},
  {Niehaus}, {Ntuli}, {Nzama}, {Obies}, {Obrocka}, {Ockards}, {Olyn}, {Oozeer},
  {Otto}, {Padayachee}, {Passmoor}, {Patel}, {Paula}, {Peens-Hough},
  {Pholoholo}, {Prozesky}, {Rakoma}, {Ramaila}, {Rammala}, {Ramudzuli},
  {Rasivhaga}, {Ratcliffe}, {Reader}, {Renil}, {Richter}, {Robyntjies},
  {Rosekrans}, {Rust}, {Salie}, {Sambu}, {Schollar}, {Schwardt}, {Seranyane},
  {Sethosa}, {Sharpe}, {Siebrits}, {Sirothia}, {Slabber}, {Smirnov}, {Smith},
  {Sofeya}, {Songqumase}, {Spann}, {Stappers}, {Steyn}, {Steyn}, {Strong},
  {Struthers}, {Stuart}, {Sunnylall}, {Swart}, {Taljaard}, {Tasse}, {Taylor},
  {Theron}, {Thondikulam}, {Thorat}, {Tiplady}, {Toruvanda}, {van Aardt}, {van
  Balla}, {van den Heever}, {van der Byl}, {van der Merwe}, {van der Merwe},
  {van Niekerk}, {van Rooyen}, {van Staden}, {van Tonder}, {van Wyk}, {Wait},
  {Walker}, {Wallace}, {Welz}, {Williams}, {Xaia}, {Young}, \&
  {Zitha}}]{Camilo18}
{Camilo}, F., {Scholz}, P., {Serylak}, M., {et~al.} 2018, \apj, 856, 180

\bibitem[{{Casasola} {et~al.}(2017){Casasola}, {Cassar{\`a}}, {Bianchi},
  {Verstocken}, {Xilouris}, {Magrini}, {Smith}, {De Looze}, {Galametz},
  {Madden}, {Baes}, {Clark}, {Davies}, {De Vis}, {Evans}, {Fritz}, {Galliano},
  {Jones}, {Mosenkov}, {Viaene}, \& {Ysard}}]{casasola2017}
{Casasola}, V., {Cassar{\`a}}, L.~P., {Bianchi}, S., {et~al.} 2017, \aap, 605,
  A18

\bibitem[{{Chengalur} {et~al.}(1999){Chengalur}, {de Bruyn}, \&
  {Narasimha}}]{chengalur1999}
{Chengalur}, J.~N., {de Bruyn}, A.~G., \& {Narasimha}, D. 1999, \aap, 343, L79

\bibitem[{{Chengalur} \& {Kanekar}(2003)}]{chengalur2003}
{Chengalur}, J.~N. \& {Kanekar}, N. 2003, \prl, 91, 241302

\bibitem[{{Chilingarian} {et~al.}(2010){Chilingarian}, {Di Matteo}, {Combes},
  {Melchior}, \& {Semelin}}]{chilingarian2010}
{Chilingarian}, I.~V., {Di Matteo}, P., {Combes}, F., {Melchior}, A.~L., \&
  {Semelin}, B. 2010, \aap, 518, A61

\bibitem[{{Combes}(2008)}]{combes2008}
{Combes}, F. 2008, arXiv e-prints, arXiv:0811.0153

\bibitem[{{Courbin} {et~al.}(1998){Courbin}, {Lidman}, {Frye}, {Magain},
  {Broadhurst}, {Pahre}, \& {Djorgovski}}]{courbin1998}
{Courbin}, F., {Lidman}, C., {Frye}, B.~L., {et~al.} 1998, \apjl, 499, L119

\bibitem[{{Courbin} {et~al.}(2002){Courbin}, {Meylan}, {Kneib}, \&
  {Lidman}}]{courbin2002}
{Courbin}, F., {Meylan}, G., {Kneib}, J.~P., \& {Lidman}, C. 2002, \apj, 575,
  95

\bibitem[{{Darling}(2003)}]{darling2003}
{Darling}, J. 2003, \prl, 91, 011301

\bibitem[{{Darling}(2004)}]{darling2004}
{Darling}, J. 2004, \apj, 612, 58

\bibitem[{{Darling} \& {Giovanelli}(2002)}]{darling2002}
{Darling}, J. \& {Giovanelli}, R. 2002, \aj, 124, 100

\bibitem[{{Elitzur}(1976)}]{Elitzur1976}
{Elitzur}, M. 1976, \apj, 203, 124

\bibitem[{{Frayer} {et~al.}(1998){Frayer}, {Seaquist}, \& {Frail}}]{frayer1998}
{Frayer}, D.~T., {Seaquist}, E.~R., \& {Frail}, D.~A. 1998, \aj, 115, 559

\bibitem[{{Frye} {et~al.}(1997){Frye}, {Welch}, \& {Broadhurst}}]{Frye1997}
{Frye}, B., {Welch}, W.~J., \& {Broadhurst}, T. 1997, \apjl, 478, L25

\bibitem[{{Garrett} {et~al.}(1997){Garrett}, {Nair}, {Porcas}, \&
  {Patnaik}}]{garrett1997}
{Garrett}, M.~A., {Nair}, S., {Porcas}, R.~W., \& {Patnaik}, A.~R. 1997, Vistas
  in Astronomy, 41, 281

\bibitem[{{Guirado} {et~al.}(1999){Guirado}, {Jones}, {Lara}, {Marcaide},
  {Preston}, {Rao}, \& {Sherwood}}]{guirado1999}
{Guirado}, J.~C., {Jones}, D.~L., {Lara}, L., {et~al.} 1999, \aap, 346, 392

\bibitem[{{Gupta} {et~al.}(2021){Gupta}, {Jagannathan}, {Srianand},
  {Bhatnagar}, {Noterdaeme}, {Combes}, {Petitjean}, {Jose}, {Pandey}, {Kaski},
  {Baker}, {Balashev}, {Boettcher}, {Chen}, {Cress}, {Dutta}, {Goedhart},
  {Heald}, {J{\'o}zsa}, {Kamau}, {Kamphuis}, {Kerp}, {Kl{\"o}ckner}, {Knowles},
  {Krishnan}, {Krogager}, {Kulkarni}, {Momjian}, {Moodley}, {Passmoor},
  {Schr{\"o}eder}, {Sekhar}, {Sikhosana}, {Wagenveld}, \& {Wong}}]{gupta2020}
{Gupta}, N., {Jagannathan}, P., {Srianand}, R., {et~al.} 2021, \apj, 907, 11

\bibitem[{{Gupta} {et~al.}(2018{\natexlab{a}}){Gupta}, {Momjian}, {Srianand},
  {Petitjean}, {Noterdaeme}, {Gyanchandani}, {Sharma}, \&
  {Kulkarni}}]{Gupta2018oh}
{Gupta}, N., {Momjian}, E., {Srianand}, R., {et~al.} 2018{\natexlab{a}}, \apjl,
  860, L22

\bibitem[{{Gupta} {et~al.}(2017){Gupta}, {Srianand}, {Baan}, {Baker},
  {Beswick}, {Bhatnagar}, {Bhattacharya}, {Bosma}, {Carilli}, {Cluver},
  {Combes}, {Cress}, {Dutta}, {Fynbo}, {Heald}, {Hilton}, {Hussain}, {Jarvis},
  {Jozsa}, {Kamphuis}, {Kembhavi}, {Kerp}, {Kl{\"o}ckner}, {Krogager},
  {Kulkarni}, {Ledoux}, {Mahabal}, {Mauch}, {Moodley}, {Momjian}, {Morganti},
  {Noterdaeme}, {Oosterloo}, {Petitjean}, {Schr{\"o}der}, {Serra}, {Sievers},
  {Spekkens}, {V{\"a}is{\"a}nen}, {van der Hulst}, {Vivek}, {Wang}, {Wong}, \&
  {Zungu}}]{Gupta17mals}
{Gupta}, N., {Srianand}, R., {Baan}, W., {et~al.} 2017, ArXiv e-prints
  [\eprint[arXiv]{1708.07371}]

\bibitem[{{Gupta} {et~al.}(2018{\natexlab{b}}){Gupta}, {Srianand}, {Farnes},
  {Pidopryhora}, {Vivek}, {Paragi}, {Noterdaeme}, {Oosterloo}, \&
  {Petitjean}}]{Gupta18j1243}
{Gupta}, N., {Srianand}, R., {Farnes}, J.~S., {et~al.} 2018{\natexlab{b}},
  \mnras, 476, 2432

\bibitem[{{Jagannathan} {et~al.}(2017){Jagannathan}, {Bhatnagar}, {Rau}, \&
  {Taylor}}]{jagannathan2017}
{Jagannathan}, P., {Bhatnagar}, S., {Rau}, U., \& {Taylor}, A.~R. 2017, \aj,
  154, 56

\bibitem[{{Jauncey} {et~al.}(1991){Jauncey}, {Reynolds}, {Tzioumis}, {Muxlow},
  {Perley}, {Murphy}, {Preston}, {King}, {Patnaik}, {Jones}, {Meier}, {Bird},
  {Blair}, {Bunton}, {Clay}, {Costa}, {Duncan}, {Ferris}, {Gough}, {Hamilton},
  {Hoard}, {Kemball}, {Kesteven}, {Lobdell}, {Luiten}, {Mcculloch}, {Murray},
  {Nicholson}, {Rao}, {Savage}, {Sinclair}, {Skjerve}, {Taaffe}, {Wark}, \&
  {White}}]{Jauncey1991}
{Jauncey}, D.~L., {Reynolds}, J.~E., {Tzioumis}, A.~K., {et~al.} 1991, \nat,
  352, 132

\bibitem[{{Jin} {et~al.}(2003){Jin}, {Garrett}, {Nair}, {Porcas}, {Patnaik}, \&
  {Nan}}]{jin2003}
{Jin}, C., {Garrett}, M.~A., {Nair}, S., {et~al.} 2003, \mnras, 340, 1309

\bibitem[{{Jonas} \& {MeerKAT Team}(2016)}]{Jonas16}
{Jonas}, J. \& {MeerKAT Team}. 2016, in Proceedings of MeerKAT Science: On the
  Pathway to the SKA. 25-27 May, 2016 Stellenbosch, South Africa (MeerKAT2016),
  1

\bibitem[{{Kanekar} \& {Chengalur}(2002)}]{kanekar2002}
{Kanekar}, N. \& {Chengalur}, J.~N. 2002, \aap, 381, L73

\bibitem[{{Kanekar} {et~al.}(2004){Kanekar}, {Chengalur}, \&
  {Ghosh}}]{kanekar2004}
{Kanekar}, N., {Chengalur}, J.~N., \& {Ghosh}, T. 2004, \prl, 93, 051302

\bibitem[{{Kanekar} {et~al.}(2018){Kanekar}, {Ghosh}, \&
  {Chengalur}}]{kanekar2018}
{Kanekar}, N., {Ghosh}, T., \& {Chengalur}, J.~N. 2018, \prl, 120, 061302

\bibitem[{{Koopmans} \& {de Bruyn}(2005)}]{koopmans2005}
{Koopmans}, L.~V.~E. \& {de Bruyn}, A.~G. 2005, \mnras, 360, L6

\bibitem[{{Leroy} {et~al.}(2009){Leroy}, {Walter}, {Bigiel}, {Usero}, {Weiss},
  {Brinks}, {de Blok}, {Kennicutt}, {Schuster}, {Kramer}, {Wiesemeyer}, \&
  {Roussel}}]{leroy2009}
{Leroy}, A.~K., {Walter}, F., {Bigiel}, F., {et~al.} 2009, \aj, 137, 4670

\bibitem[{{Li} {et~al.}(2018){Li}, {Tang}, {Nguyen}, {Dawson}, {Heiles}, {Xu},
  {Pan}, {Goldsmith}, {Gibson}, {Murray}, {Robishaw}, {McClure-Griffiths},
  {Dickey}, {Pineda}, {Stanimirovi{\'c}}, {Bronfman}, {Troland}, \& {PRIMO
  Collaboration}}]{Li2018}
{Li}, D., {Tang}, N., {Nguyen}, H., {et~al.} 2018, \apjs, 235, 1

\bibitem[{{Lidman} {et~al.}(1999){Lidman}, {Courbin}, {Meylan}, {Broadhurst},
  {Frye}, \& {Welch}}]{lidman1999}
{Lidman}, C., {Courbin}, F., {Meylan}, G., {et~al.} 1999, \apjl, 514, L57

\bibitem[{{Liszt} \& {Lucas}(1996)}]{Liszt96}
{Liszt}, H. \& {Lucas}, R. 1996, \aap, 314, 917

\bibitem[{{Lovell} {et~al.}(1998){Lovell}, {Jauncey}, {Reynolds}, {Wieringa},
  {King}, {Tzioumis}, {McCulloch}, \& {Edwards}}]{lovell1998}
{Lovell}, J.~E.~J., {Jauncey}, D.~L., {Reynolds}, J.~E., {et~al.} 1998, \apjl,
  508, L51

\bibitem[{{Lovell} {et~al.}(1996){Lovell}, {Reynolds}, {Jauncey}, {Backus},
  {McCulloch}, {Sinclair}, {Wilson}, {Tzioumis}, {King}, {Gough}, {Ellingsen},
  {Phillips}, {Preston}, \& {Jones}}]{lovell1996}
{Lovell}, J.~E.~J., {Reynolds}, J.~E., {Jauncey}, D.~L., {et~al.} 1996, \apjl,
  472, L5

\bibitem[{{Marasco} {et~al.}(2019){Marasco}, {Fraternali}, {Heald}, {de Blok},
  {Oosterloo}, {Kamphuis}, {J{\'o}zsa}, {Vargas}, {Winkel}, {Walterbos},
  {Dettmar}, \& {Juẗte}}]{marasco2019}
{Marasco}, A., {Fraternali}, F., {Heald}, G., {et~al.} 2019, \aap, 631, A50

\bibitem[{{Mart{\'\i}-Vidal} {et~al.}(2013){Mart{\'\i}-Vidal}, {Muller},
  {Combes}, {Aalto}, {Beelen}, {Darling}, {Gu{\'e}lin}, {Henkel}, {Horellou},
  {Marcaide}, {Mart{\'\i}n}, {Menten}, {V-Trung}, \& {Zwaan}}]{marti2013}
{Mart{\'\i}-Vidal}, I., {Muller}, S., {Combes}, F., {et~al.} 2013, \aap, 558,
  A123

\bibitem[{{Mauch} {et~al.}(2020){Mauch}, {Cotton}, {Condon}, {Matthews},
  {Abbott}, {Adam}, {Aldera}, {Asad}, {Bauermeister}, {Bennett}, {Bester},
  {Botha}, {Brederode}, {Brits}, {Buchner}, {Burger}, {Camilo}, {Chalmers},
  {Cheetham}, {de Villiers}, {de Villiers}, {Dikgale-Mahlakoana}, {du Toit},
  {Esterhuyse}, {Fadana}, {Fanaroff}, {Fataar}, {February}, {Frank},
  {Gamatham}, {Geyer}, {Goedhart}, {Gounden}, {Gumede}, {Heywood}, {Hlakola},
  {Horrell}, {Hugo}, {Isaacson}, {J{\'o}zsa}, {Jonas}, {Julie}, {Kapp},
  {Kasper}, {Kenyon}, {Kotz{\'e}}, {Kriek}, {Kriel}, {Kusel}, {Lehmensiek},
  {Loots}, {Lord}, {Lunsky}, {Madisa}, {Magnus}, {Main}, {Malan}, {Manley},
  {Marais}, {Martens}, {Merry}, {Millenaar}, {Mnyandu}, {Moeng}, {Mokone},
  {Monama}, {Mphego}, {New}, {Ngcebetsha}, {Ngoasheng}, {Ockards}, {Oozeer},
  {Otto}, {Patel}, {Peens-Hough}, {Perkins}, {Ramaila}, {Ramudzuli}, {Renil},
  {Richter}, {Robyntjies}, {Salie}, {Schollar}, {Schwardt}, {Serylak},
  {Siebrits}, {Sirothia}, {Smirnov}, {Sofeya}, {Stone}, {Taljaard}, {Tasse},
  {Theron}, {Tiplady}, {Toruvanda}, {Twum}, {van Balla}, {van der Byl}, {van
  der Merwe}, {Van Tonder}, {Wallace}, {Welz}, {Williams}, \& {Xaia}}]{Mauch20}
{Mauch}, T., {Cotton}, W.~D., {Condon}, J.~J., {et~al.} 2020, \apj, 888, 61

\bibitem[{{Mohan} \& {Rafferty}(2015)}]{mohan2015}
{Mohan}, N. \& {Rafferty}, D. 2015, {PyBDSF: Python Blob Detection and Source
  Finder}

\bibitem[{{M{\"u}ller} {et~al.}(2015){M{\"u}ller}, {Muller}, {Schilke},
  {Bergin}, {Black}, {Gerin}, {Lis}, {Neufeld}, \& {Suri}}]{muller2015}
{M{\"u}ller}, H. S.~P., {Muller}, S., {Schilke}, P., {et~al.} 2015, \aap, 582,
  L4

\bibitem[{{Muller} {et~al.}(2011){Muller}, {Beelen}, {Gu{\'e}lin}, {Aalto},
  {Black}, {Combes}, {Curran}, {Theule}, \& {Longmore}}]{muller2011}
{Muller}, S., {Beelen}, A., {Gu{\'e}lin}, M., {et~al.} 2011, \aap, 535, A103

\bibitem[{{Muller} {et~al.}(2014){Muller}, {Combes}, {Gu{\'e}lin}, {G{\'e}rin},
  {Aalto}, {Beelen}, {Black}, {Curran}, {Darling}, {V-Trung},
  {Garc{\'\i}a-Burillo}, {Henkel}, {Horellou}, {Mart{\'\i}n},
  {Mart{\'\i}-Vidal}, {Menten}, {Murphy}, {Ott}, {Wiklind}, \&
  {Zwaan}}]{muller2014}
{Muller}, S., {Combes}, F., {Gu{\'e}lin}, M., {et~al.} 2014, \aap, 566, A112

\bibitem[{{Muller} \& {Gu{\'e}lin}(2008)}]{muller2008}
{Muller}, S. \& {Gu{\'e}lin}, M. 2008, \aap, 491, 739

\bibitem[{{Muller} {et~al.}(2020){Muller}, {Jaswanth}, {Horellou}, \&
  {Mart{\'\i}-Vidal}}]{muller2020}
{Muller}, S., {Jaswanth}, S., {Horellou}, C., \& {Mart{\'\i}-Vidal}, I. 2020,
  \aap, 641, L2

\bibitem[{{Muller} {et~al.}(2017){Muller}, {M{\"u}ller}, {Black}, {G{\'e}rin},
  {Combes}, {Curran}, {Falgarone}, {Gu{\'e}lin}, {Henkel}, {Mart{\'\i}n},
  {Menten}, {Roueff}, {Aalto}, {Beelen}, {Wiklind}, \& {Zwaan}}]{muller2017}
{Muller}, S., {M{\"u}ller}, H.~S.~P., {Black}, J.~H., {et~al.} 2017, \aap, 606,
  A109

\bibitem[{{Nair} {et~al.}(2005){Nair}, {Jin}, \& {Garrett}}]{nair2005}
{Nair}, S., {Jin}, C., \& {Garrett}, M.~A. 2005, \mnras, 362, 1157

\bibitem[{{Nair} {et~al.}(1993){Nair}, {Narasimha}, \& {Rao}}]{nair1993}
{Nair}, S., {Narasimha}, D., \& {Rao}, A.~P. 1993, \apj, 407, 46

\bibitem[{{Olling}(1996)}]{olling1996}
{Olling}, R.~P. 1996, \aj, 112, 457

\bibitem[{{Oshima} {et~al.}(2001){Oshima}, {Mitsuda}, {Ota}, {Yonehara},
  {Hattori}, {Mihara}, \& {Sekimoto}}]{oshima2001}
{Oshima}, T., {Mitsuda}, K., {Ota}, N., {et~al.} 2001, \apj, 551, 929

\bibitem[{{Patnaik} {et~al.}(1993){Patnaik}, {Muxlow}, \&
  {Jauncey}}]{patnaik1993}
{Patnaik}, A.~R., {Muxlow}, T.~W.~B., \& {Jauncey}, D.~L. 1993, in Liege
  International Astrophysical Colloquia, Vol.~31, Liege International
  Astrophysical Colloquia, ed. J.~{Surdej}, D.~{Fraipont-Caro}, E.~{Gosset},
  S.~{Refsdal}, \& M.~{Remy}, 363

\bibitem[{{Pramesh Rao} \& {Subrahmanyan}(1988)}]{Rao1988}
{Pramesh Rao}, A. \& {Subrahmanyan}, R. 1988, \mnras, 231, 229

\bibitem[{{Rau} \& {Cornwell}(2011)}]{rau2011}
{Rau}, U. \& {Cornwell}, T.~J. 2011, \aap, 532, A71

\bibitem[{{Srianand} {et~al.}(2013){Srianand}, {Gupta}, {Rahmani}, {Momjian},
  {Petitjean}, \& {Noterdaeme}}]{Srianand13dib}
{Srianand}, R., {Gupta}, N., {Rahmani}, H., {et~al.} 2013, \mnras, 428, 2198

\bibitem[{{Tercero} {et~al.}(2020){Tercero}, {Cernicharo}, {Cuadrado}, {de
  Vicente}, \& {Gu{\'e}lin}}]{tercero2020}
{Tercero}, B., {Cernicharo}, J., {Cuadrado}, S., {de Vicente}, P., \&
  {Gu{\'e}lin}, M. 2020, \aap, 636, L7

\bibitem[{{van Langevelde} {et~al.}(1995){van Langevelde}, {van Dishoeck},
  {Sevenster}, \& {Israel}}]{langevelde1995}
{van Langevelde}, H.~J., {van Dishoeck}, E.~F., {Sevenster}, M.~N., \&
  {Israel}, F.~P. 1995, \apjl, 448, L123

\bibitem[{{Walter} {et~al.}(2008){Walter}, {Brinks}, {de Blok}, {Bigiel},
  {Kennicutt}, {Thornley}, \& {Leroy}}]{walter2008}
{Walter}, F., {Brinks}, E., {de Blok}, W.~J.~G., {et~al.} 2008, \aj, 136, 2563

\bibitem[{{Wiklind} \& {Combes}(1996)}]{wiklind1996}
{Wiklind}, T. \& {Combes}, F. 1996, \nat, 379, 139

\bibitem[{{Wiklind} \& {Combes}(1998)}]{wiklind1998}
{Wiklind}, T. \& {Combes}, F. 1998, \apj, 500, 129

\bibitem[{{Wiklind} \& {Combes}(2001)}]{wiklind2001}
{Wiklind}, T. \& {Combes}, F. 2001, in Astronomical Society of the Pacific
  Conference Series, Vol. 237, Gravitational Lensing: Recent Progress and
  Future Go, ed. T.~G. {Brainerd} \& C.~S. {Kochanek}, 155

\bibitem[{{Winn} {et~al.}(2002){Winn}, {Kochanek}, {McLeod}, {Falco}, {Impey},
  \& {Rix}}]{winn2002}
{Winn}, J.~N., {Kochanek}, C.~S., {McLeod}, B.~A., {et~al.} 2002, \apj, 575,
  103

\bibitem[{{Wolfe} {et~al.}(2008){Wolfe}, {Jorgenson}, {Robishaw}, {Heiles}, \&
  {Prochaska}}]{Wolfe2008}
{Wolfe}, A.~M., {Jorgenson}, R.~A., {Robishaw}, T., {Heiles}, C., \&
  {Prochaska}, J.~X. 2008, \nat, 455, 638

\end{thebibliography}
 
 \begin{appendix}
 \section{Continuum maps} \label{app:contmaps}
\begin{figure}
\includegraphics[width=0.48\textwidth,angle=0]{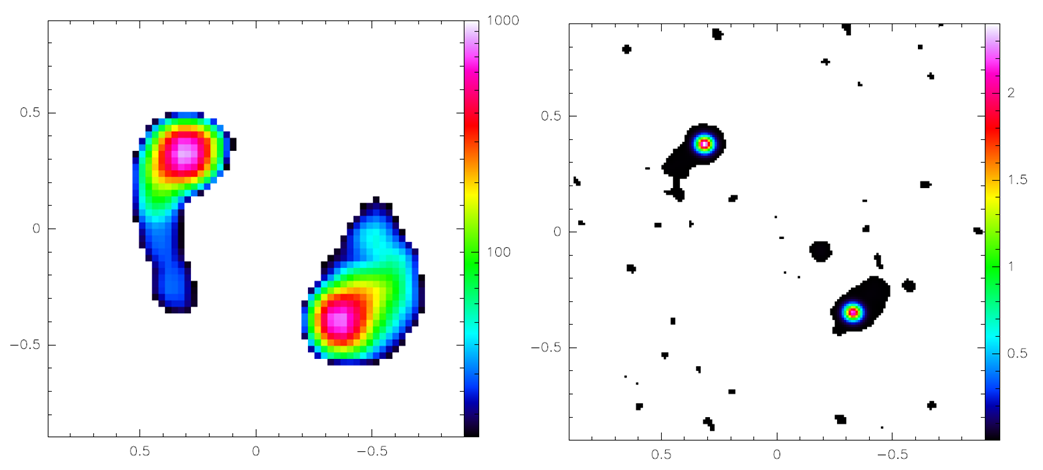}  
\vskip+0.0cm  
\caption{Continuum maps of \pks\,  used in the kinematical models:
{\bf Left:} Continuum emission at 5~GHz with synthesized beam of 0.1 \arcsec (pixel size = 0.028\,\arcsec) measured by the MERLIN interferometer
\citep{patnaik1993}, used for the \hi\ and OH spectra (color wedge in mJy/beam). {\bf Right:} Continuum emission in Band 5 ($\sim$~180GHz) of ALMA \citep{muller2020}, with beam 0.056 \arcsec (pixel size = 0.014\,\arcsec), used for HCO$^+$ and H$_2$O lines (color wedge in Jy/beam). The axes are labelled in arcsec.
} 
\label{fig:pks-cont}   
\end{figure}

 To model the continuum at cm and mm wavelengths, we adopted the MERLIN 5~GHz image from \citet{patnaik1993},
 and the ALMA-Band~5 ($\sim 180$\,GHz)
 image from \citet{muller2020} respectively, which are plotted in Fig.\ref{fig:pks-cont}. The resulting absorption lines depend essentially on the continuum distribution and not their absolute values. There might be some variations with frequency, with a
 spectral index, which is not quite flat (see NED and Section~\ref{sec:obs}), but these are of second order here.

 %%%%%%%%%%%%%%%%%%%%%%%%%%%%%
%%%%%%%%%%%%%%%%%%%%%%%%%%%%%

\section{Spectra due to the different continuum components}

\begin{figure}
\includegraphics[width=0.48\textwidth,angle=0]{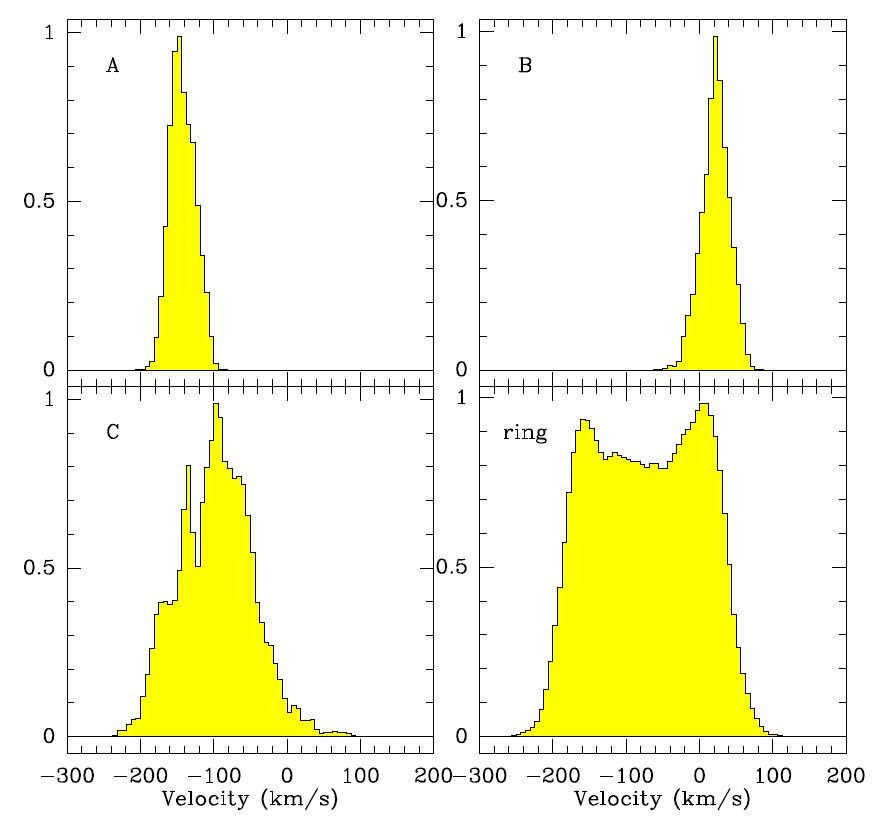}  
\vskip+0.0cm  
\caption{Spectral contributions of the different continuum sources:
{\bf Top:} the main quasar core images, A(NE) and B(SW); {\bf Bottom:} the third image C see Fig. \ref{fig:galaxy-model}, and the Einstein ring, modeled as a constant continuum flux in the elliptical ring, schematically drawn in Fig. \ref{fig:galaxy-model}. 
} 
\label{fig:fig-mod}   
\end{figure}

To better understand relative contributions to absorption from different continuum images of the background quasar, we decompose the spectra in front of the following image components: the three images of the quasar core, and the Einstein ring (see Fig. \ref{fig:fig-mod}). All spectra in the figure are normalized to 1.

 %%%%%%%%%%%%%%%%%%%%%%%%%%%%%
%%%%%%%%%%%%%%%%%%%%%%%%%%%%%

 \end{appendix}
 
\end{document}